\documentclass[UTF8,a4paper]{article}
\usepackage{booktabs}
\usepackage{amsmath}
\usepackage{graphicx}
\usepackage{subfigure}
\usepackage{epstopdf}
\usepackage{geometry}
\usepackage{ulem}
\usepackage{indentfirst}
\usepackage{amsfonts,amssymb,dsfont}
\usepackage{setspace}
\usepackage{threeparttable}
\usepackage{amsmath,mathtools,amsthm}
\usepackage{array}
\usepackage{extarrows}
\usepackage{upgreek}

\makeatletter
\renewcommand*\env@matrix[1][\arraystretch]{%
  \edef\arraystretch{#1}%
  \hskip -\arraycolsep
  \let\@ifnextchar\new@ifnextchar
  \array{*\c@MaxMatrixCols c}}
\makeatother
\begin{document}


\title{\bf Effects of the symmetries and related orders to the thermalization of many-body localized system}
\author{Chen-Huan Wu
\thanks{chenhuanwu1@gmail.com}
\\Songshan Lake Materials Laboratory, Dongguan, Guangdong 523808, China}

\maketitle
\vspace{-30pt}
\begin{abstract}
\begin{large} 
In this paper, we discuss the effects of the symmetries and related topological
orders to the thermalization of many-body localized system.
We consider the one-dimensional fermion chain system with open boundary condition,
whose boundaries are characterized by Sachdev-Ye-Kitaev (SYK) intercation.
Just like in the SYK model and the tendor models,
there are many-body quantum chaos in out-of-time-ordered correlation
when the system is being thermalized by the interactions (usually nonuniform and being randomly distributed),
and satisfies the eigenstate thermalization hypothesis (ETH).
While the continuous or discrete symmetries usually protect the related topological orders against the thermalization,
which
may lead to the localization of quantum states and generate large degeneracy.
We discuss these effects in terms of the fermionic or spin languages.
In many-body localized phase, a large number of degenerate 
ETH-violated states can be found in both the frustration-free 
AKLT model or the integrable biquadratic (Wishart) SYK model.
The quantum scars appear when such degenerate states are embeded into an ETH-satisfying spectrum
of the Hilbert space enlarged by a much larger number of bosonic flavors (e.g., SU(M) multiplets degeneracy).
Usually, the emergence of quantum chaos require large limits of boson flavor $M$ and the number of coupled (undegenerate) states ($N$;
which related to $\mathds{Z}_{N}$ symmetry).
Further, the boson (or excitations) flavor number $M$ can often to related to the fermion number $N$ through the duality transformations,
in which case the $\mathds{Z}_{M}$ symmetry is possible to generted by SU(M) model and leads to asymptotic degeneracy in large-$M$ limit.
\\

\end{large}

\end{abstract}
\begin{large}

\section{Introduction}

The SYK physics emerges usually in the non-Fermi liquid phase with SYK coupling much larger than the fermionic frequency and while the 
fermionic frequency much larger than the coherence scala (incoherent critical metal phase).
In such a regime, the quantum critical behaviors can be found in itinerant electron system due to the quantum fluctuation induced
quantum phase transition between disordered state and the Fermi liquid state.
In a many-body system with disorder-induced random interactions controlled by SYK physics,
the many-body energy spectrum is consistituted by the continuous distributed energy levels with finite energy spacings,
and as long as the system is in the non-fermi liquid phase,
the level spacing is a local property and depends only on the symmetry class of the Hamiltonian matrix.
The signature for the thermalization of many-body localized symmetry-protected topological states 
is the emergence of global quantum anomaly in $\mathds{Z}$ classfication,
and the resulting eigenstate thermalization hypothsis (ETH) is similar to the effect of unitary evolution in
 random quantum circuit which leads to volume-law scaling entanglement entropies as long as the random local measurement
is lower than the critical value.
The standard SYK non-fermi liquid also follows the ETH,
with the level statistics follow the Wigner-Dyson distribution,
while the bilinear term (or fermion number operator which is conserved mod 4)
can leads to spontaneous symmetry breaking and exponential decay of long imaginary time correlation which is formed in thermal equilibrium.
Here the perturbation from bilinear term is similar to that from symmetry-breaking order or topological
order in many-body localized system (MBL) with strong disorder, and the local measurement in a quantum circuit\cite{Nahum A}.
We will focus on the extensive thermodynamic entropy regime in this paper, which satisfies a volume-law scaling,
and the large size of bipartite system $L$ corresponds to the large $N$ limit in the replica trick approach.
The thermalized states are usually the highly excited states,
whose long time behavior is described by equilibrium statistical mechanics, insteads of the quantum ground states.
Here the long time behavior corresponds to the long imaginary time correlation in SYK model,
i.e., the universal propagator between two excited states (replica) belonging to the same symmetry sector,
and the ground state energy can be determined by this propagator.

In this paper, we focus on the one-dimensional (1D) fermion chain model,
with symmetry-protected topological (SPT) orders (both the trivial and nontrivial ones)
origin from the disorder of MBL bulk
and disorder-induced randomness.
A typical example is the zigzag silicene nanoribbon, where the chiral symmetry exists due to the intrinsic pseudospin degree-of-freedom,
unlike the graphene which requires a disorder acting on the Dirac point to creates the chiral symmetry.
Specially, 
the Majorana zero mode (MZM) exists in the edge states of the Kitaev chain,
and the number of MZMs at the boundary $N_{\chi}$ can be recognized as an integer topological index.
Experimentally, a MZM can be realized in the Kitaev chain or by coupling the Dirac fermions to a soliton-like structure like the vortices.
We will known, through the following discussions, that for the model of Majorana chains in different phases,
the fermion bilinear term $i\chi_{a}\chi_{b}$ is forbiddened by the (antiunitary) time-reversal symmetry,
or other $\mathds{Z}_{n}$ symmetries where $n$ is an even number,
when $a$ and $b$ belong to the same fermion parity (all even or all odd).
While when $a$ and $b$ belong to different fermion parities,
although the two-fold degenerated state $i\chi_{a}\chi_{b}$ satisfies time-reversal symmetry,
it can be gapped out by boundary Hamiltonian as long as the interactions exist.
Thus the fermion bilinear term is impossible to exits unless with zero interaction.
In the special case that the symmetry $\mathds{Z}_{n}$ exists but with zero interaction,
the free fermion system will exihibts nontrivial topological orders,
e.g., the nontrivial topological superconductivity.
Thus the boundaries of the chains can only be governed, to lowest order, by the SYK$_{4}$ dynamics,
due to the random Gaussian-distributed interaction acting on the boundary.
The emergent SYK physics here also corresponds to the thermalization and the unitary evolution,
where the unitary classes can be further discussed.
The SYK Hamiltonian describing the boundary states reads
\begin{equation} 
\begin{aligned}
H=\sum_{ijkl}^{N}J_{ijkl}\chi_{i}\chi_{j}\chi_{k}\chi_{l},
\end{aligned}
\end{equation}
where $J_{ijkl}$ is the random all-to-all interaction with zero mean $\overline{J_{ijkl}}=0$,
and variance $\overline{J_{ijkl}^{2}}=\frac{J^{2}}{4N^{3}}$.
And the interaction makes the classification of $\mathds{Z}$ invariant be broken down to the one with multi-fold periodicity,
i.e., with global anomaly.
So the $\mathds{Z}$ symmetry can be preserved no matter how much the MZM is,
only when the interaction is absent.
$H$ here can be viewed as a matrix in Hilbert (Fock) space with dimension exponential in $N$ 
and depends on the eigenvalue of U(1) charge operator.
During the discusion of the related topological invariant under certain operations,
it is sometimes more convenient to make the transformation of Fock basis
between fermion operator to the Majorana operator.

We note that, the above boundary Hamiltonian can describes both the SYK mode
in the boundaries of 1D Kitaev chain and the 1D Majorana chain.
This is base on the conclusion of Ref.\cite{Fendley P}:
exact edge zero modes can exist in both Ising chain and Majorana chain,
and in both cases the topologically order is possible.
We will focus mostly on the topological order/disorder of Majorana chain in this paper,
where the thermalization is absent in the bulk part due to the existence of local conserved quantities,
while in the boundaries,
although the conservation may remain (like the U(1) gauge symmetry in charge or spin channel),
it is not local any more due to the generic (without integrability) nonlocal and dense SYK interaction.
The states localized or thermalized will exhibit different properties,
including the level statistic, spectral density and long-time correlation (in terms of fermions or spins).

In Eq.(1),
the Gaussian-distributed $J_{ijkl}$ breaks the integrability since the interaction $J_{ijkl}$ if of the random-range type.
The absence of integrability is the precondition to make sure all eigenstates of the ergodic system described by above Hamiltonian
follow ETH.
Any long-range hopping, long-range interaction, and on-site disorder (i.e., the on-site interaction with a site index-dependent strength)
will break the integrability of 1D system.
Thus the integrable Majorana chain model can exist only in the absence of $J_{ijkl}$,
like the simplest example, which process a global $\mathds{Z}_{2}$ symmetry:
$H=-\sum^{(N_{\chi}-2)/2}_{i=1}iA_{i}\chi_{2i-1}\chi_{2i+2}-\sum^{N_{\chi}/2}_{j=1}iB_{j}\chi_{2j}\chi_{2j-1}$ ($A_{i}$ and $B_{j}$ are not Gaussian variables),
whose integarbility can be seen using the the Bethe ansatz after Jordan-Wigner transfromation:
$H=-\sum^{(N_{\chi}-2)/2}_{i=1}A_{i}\sigma^{x}_{i}\sigma^{x}_{i+1}+\sum^{N_{\chi}/2}_{j=1}B_{j}\sigma_{j}^{z}$.

As the solvable SYK model requires a $N_{\chi}\rightarrow\infty$ limit,
we will apply the method of matrix product state (MPS) insteads of the exact diagonalization.
Since the main disorder we consider is the one induced by MBL in the bulk,
the matrix Hamiltonian should be block-diagonalized into sectors operated by unitary symmetries.
While if the main disorder is of the random-hopping character,
the Hamiltonian should be block off-diagolized 
as the ramdom hopping (or the random on-site potential) is of the off-diagonal disorder.
The case dominated by this kind of disorder is being studied in Ref.\cite{Konig E J}.

\section{CII}

The CII class corresponds to the preserved time-reversal symmetry (TRS) and broken spin rotation symmetry (SRS),
due to the nonlocal feature of order paramater in terms of spin.
Since the SYK system is zero-dimensional (0D) in spacial space, and usually does not contains any momentum component 
(the momentum component can be added in some cases),
the TRS can be written as $\mathcal{T}H\mathcal{T}^{-1}=H$,
with the Hamiltonian $H$ has being diagonalized and thus $H=H^{T}$.
Then we introduce the antiunitary chiral symmetry operator $\mathcal{S}=\mathcal{T}\mathcal{C}\mathcal{K}$.
Here we note that,
$\mathcal{C}$ is the particle-hole symmetry (PHS) operator,
which reads $\mathcal{C}=\prod^{N}_{i}(c_{i}+c_{i}^{\dag})$ in many-body system and $\mathcal{K}$ is the complex conjugation operator
which equivlents to spinless antiunitary time reversal
(since complex conjugation operator is involutory, we have $\mathcal{K}=\mathcal{K}^{-1}$),
and they satisfy
\begin{equation} 
\begin{aligned}
&\mathcal{C}^{-1}c_{\uparrow}\mathcal{C}=c_{\downarrow}^{\dag},\\
&\mathcal{C}^{-1}c_{\downarrow}\mathcal{C}=-c_{\uparrow}^{\dag},\\
&\mathcal{C}^{-1}c_{\uparrow}^{\dag}\mathcal{C}=c_{\downarrow},\\
&\mathcal{C}^{-1}c_{\downarrow}^{\dag}\mathcal{C}=-c_{\uparrow},\\
&\mathcal{C}^{-1}\mathcal{Q}^{*}\mathcal{C}=-\mathcal{Q},\\
&\mathcal{C}^{-1}H^{*}\mathcal{C}=H,\\
&\mathcal{K}c^{\dag}\mathcal{K}^{-1}=c,\\
&\mathcal{K}c\mathcal{K}^{-1}=c^{\dag},
\end{aligned}
\end{equation}
where $\mathcal{Q}=\sum^{N}_{i}(c^{\dag}_{i}c_{i}-1/2)$ is the U(1) globally conserved charge operator,
with eigenvalues $\lambda_{Q}=\pm \sum^{N}_{i}(n_{i}-1/2)$ ($n_{i}$ is the fermion number operator and thus charge neutrality
$\lambda_{Q}=0$ corresponds to $N/2$ particle).
This U(1) charge is equivalent to the fermion number operator in the Fock basis of Majorana operator 
in the trivial phase (in the absence of twofold ground-state degeneracy).
Thus we can further know, in this class,
$\mathcal{C}^{2}=(-1)^{-\lambda_{Q}}$,
and $\mathcal{S}^{2}=(-1)^{N}$.

Then the TRS in CII class can be expressed by
\begin{equation} 
\begin{aligned}
\mathcal{T}H\mathcal{T}^{-1}=&\mathcal{S}\mathcal{C}^{-1}\mathcal{K}^{-1}H\mathcal{K}\mathcal{C}\mathcal{S}^{-1}\\
=&\mathcal{S}\mathcal{C}^{-1}H^{*}\mathcal{C}\mathcal{S}^{-1}\\
=&\mathcal{S}H\mathcal{S}^{-1}\\
=&H.
\end{aligned}
\end{equation}
So we have $\mathcal{T}^{2}=\mathcal{S}^{2}\mathcal{C}^{-2}=(-1)^{N+\lambda_{Q}}$.
It is also easy to see that, the TRS (as well as the spatial parity symmetry, like pseudospin) is broken in the absence of PHS,
which is consistent with the effect of chiral interactions\cite{Fendley P}.
This is easy to verified since $\mathcal{T}^{2}$ is closely related to the $\mathds{Z}_{2}$ fermionic number parity 
which is given by
$\mathcal{Q}_{s}=2S_{z}$ in the Majorana basis with trivial phase,
where $\chi_{i}$ is the Majorana fermion operator with index denotes their position in the chain.
By impose TRS operator to the fermion operator (in real representation),
we have
\begin{equation} 
\begin{aligned}
\mathcal{T}c_{i}\mathcal{T}^{-1}=&\mathcal{S}\mathcal{C}^{-1}\mathcal{K}^{-1}c_{i}\mathcal{K}\mathcal{C}\mathcal{S}^{-1}\\
=&\mathcal{S}\mathcal{C}^{-1}c_{i}\mathcal{C}\mathcal{S}^{-1}\\
=&\mathcal{S}c_{i}^{\dag}\mathcal{S}^{-1}\\
=&c_{i},
\end{aligned}
\end{equation}
which just like the complex conjugation operation.

It is also easy to verify that in this class the $H$ does not satisfy the antiunitary TRS,
since $\mathcal{T}\mathcal{K}H\mathcal{K}^{-1}\mathcal{T}^{-1}=\mathcal{S}H^{*}\mathcal{S}^{-1}=H^{*}$,
unless $H$ contains only real or real quaterinion entries.
However, such an antiunitary TRS is preserved in the system of two coupled complex SYK models (ccSYK),
where the spin component of the fermion operators are considered.

Since the antiunitary chiral symmetry can be decomposed into the unitary matrix and complex conjugation operator,
$\mathcal{S}=\mathcal{T}\mathcal{C}\mathcal{K}=\mathcal{U}_{CII}\mathcal{K}$,
we can then define the unitary matrix in CII class as $\mathcal{U}_{CII}=\mathcal{T}\mathcal{C}$.
It can also be verified that,
unlike the case of single-particle level\cite{You Y Z},
the unitary matrix here obeys 
\begin{equation} 
\begin{aligned}
\mathcal{U}^{T}_{CII}
=&(\mathcal{T}\mathcal{C})^{T}\\
=&\mathcal{C}^{T}\mathcal{T}^{T}\\
=&\begin{pmatrix}0 & \mathcal{C}^{2}\\1 & 0\end{pmatrix}^{T}
\begin{pmatrix}0 & 1\\\mathcal{T}^{2} & 0\end{pmatrix}^{T}\\
=&(-1)^{N+\lambda_{Q}}\mathcal{C}\mathcal{T}\\
=&(-1)^{N+\lambda_{Q}}\mathcal{T}\mathcal{C}.
\end{aligned}
\end{equation}
This is can be treated as the linear constraint in CII class.
While in the absence of fermion number parity, $\mathcal{U}_{CII}$ must be a symmetry matrix,
with $\mathcal{U}_{CII}=\mathcal{U}_{CII}^{T}$.

Since CII class contains the spin-orbit coupling effect,
we treating the conserved $z$-direction spin as a U(1) charge operator $\mathcal{Q}_{s}=2S_{z}$.
The conserved spin $\mathcal{Q}_{s}$ is used here insteads of the real U(1) charge symmetry $\mathcal{Q}$ because
in the presence of $Z_{2}$ symmetry the fermion number parity is broken, and
thus $\mathcal{Q}$ is no more conserved.
This will be discussed in detail in following sections.
When the SYK Hamiltonian is being full antisymmetrized,
it is particle-hole symmetry at zero chemical potential (where $\mu=0$ and thus $\mathcal{Q}=0$),
in this case,
the system has an antiunitary PHS symmetry,
which is defined as 
$\mathcal{P}=\mathcal{C}\mathcal{K}=\prod^{N}_{i}\prod_{\sigma=\uparrow,\downarrow}(c_{i\sigma}+c_{i\sigma}^{\dag})\mathcal{K}$,
\begin{equation} 
\begin{aligned}
\mathcal{P}H\mathcal{P}^{-1}
=&\mathcal{C}\mathcal{K}H\mathcal{K}^{-1}\mathcal{C}^{-1}\\
=&\mathcal{C}H^{*}\mathcal{C}^{-1}\\
=&H,
\end{aligned}
\end{equation}
with $\mathcal{P}^{2}=\mathcal{C}^{2}=(-1)^{-\lambda_{Q}}$.

Although the spin rotation symmetry is broken in CII class,
the spin in $z$ direction $S_{z}$ could still be conserved,
which can be treated as a U(1) sector.
In this case, we can define charge conjugation operator $\mathcal{C}'=\mathcal{K}^{-1}\mathcal{C}^{-1}$,
which has $\mathcal{C}'^{2}=(-1)^{\lambda_{Q}}$.
Note that the charge conjugation operator in CII class has the following properties
\begin{equation} 
\begin{aligned}
&\mathcal{C}^{'}c_{\uparrow}\mathcal{C}^{'\dag}=c_{\downarrow},\\
&\mathcal{C}^{'}c_{\downarrow}\mathcal{C}^{'\dag}=-c_{\uparrow},\\
&\mathcal{C}^{'}c_{\uparrow}^{\dag}\mathcal{C}^{'\dag}=c_{\downarrow}^{\dag},\\
&\mathcal{C}^{'}c_{\downarrow}^{\dag}\mathcal{C}^{'\dag}=-c_{\uparrow}^{\dag},\\
&\mathcal{C}^{'}\mathcal{Q}\mathcal{C}^{'\dag}=-\mathcal{Q},\\
&\mathcal{C}^{'}H\mathcal{C}^{'\dag}=H.
\end{aligned}
\end{equation}

In the presence of block diagonalization in each U(1) charge sector,
we have the relation
\begin{equation} 
\begin{aligned}
&\mathcal{P}\mathcal{Q}\mathcal{P}^{-1}
=&\mathcal{C}\mathcal{K}\mathcal{Q}\mathcal{K}^{-1}\mathcal{C}^{-1}\\
=&\mathcal{C}\mathcal{Q}^{*}\mathcal{C}^{-1}\\
=&-\mathcal{Q}.
\end{aligned}
\end{equation}
Here $\mathcal{P}^{2}=(-1)^{N}$ when the level statistic is collected in $\mathds{Z}_{2}$ class,
and $\mathcal{P}^{2}=(-1)^{N(N-1)/2}$ when the level statistic is collected in $\mathds{Z}_{4}$ class.
Once the full antisymmetrization is done for the SYK Hamiltonian,
there will be a two-fold periodicity in fermion quantum number $N$ for  $\mathds{Z}_{2}$ model and four-fold periodicity for $\mathds{Z}_{4}$ model.

\section{AIII}

While in the unitary class AIII,
since the time-reversal symmetry is broken,
$\mathcal{P}$ is similar to the chiral symmetry, with $\mathcal{P}^{2}=(-1)^{N(N-1)/2}$.
In this case, when $H$ preserves the fermion parity ($N$ mod 4) or charge parity ($\mathcal{Q}$ mod 2),
there are protected two-fold degeneracy by the PHS with discrete anomaly in a discrete spectrum;
while when $H$ does not preserve fermion parity ($N$ mod 4) or charge parity ($\mathcal{Q}$ mod 2),
the $\mathds{Z}_{2}$ symmetry (like the TRS and PHS) as well as the other discrete symmetries will be broken in the large $N$ limit
 by the asymptotic degeneracy. 
The antiunitary TRS (like in the ccSYK) and the some $\mathds{Z}_{4}$ symmetries are free from this symmetry-breaking effect.
In this case the fermion number operator or the pairing order parameter will
require a finite (anomalous) expectation value,
which will leads to exponential decay of long-time correlation and a gap in many-body spectrum,
thus suppresses the thermalization in SYK non-fermi liquid phase and leads to an non-chaotic state with zero ground state entropy.
Note that here the expectation value should be the many-body ground state or thermal expectation value,
and large-N will leads to mean-field solution which signaling an instability to the conformal phase.

For $\mathds{Z}_{4}$ model,
as long as the Hamiltonian is being block diagonalized in U(1) symmetry sectors,
the two-fold Kramers degeneracy can only happen at charge neutrality with $N{\rm mod}4=2$.
This symmetry-protected degeneracy also reflects the quantum anomaly.
When $N$ is divisible by 4, the $\mathds{Z}_{2}$ invariant is missing,


\section{BDI}

We note that, while in chiral orthogonal class (BDI)
which describes the eight-fold anomaly with $\mathds{Z}_{8}$ symmetry in interacting system,
this linear constraint becomes $\mathcal{U}_{BDI}=\sigma_{y}\mathcal{U}_{BDI}^{T}\sigma_{y}$.
In this case $\mathcal{U}_{BDI}$ is an identity matrix with $\mathcal{U}_{BDI}^{2}=\mathcal{U}_{BDI}\mathcal{U}_{BDI}^{T}=1$.
In BDI, we also have $\mathcal{T}^{2}=\mathcal{S}^{2}=1$,
and it is more stable against interaction comparing to the CII and AIII
due to the absence of extensive degeneracy,
which makes the spontaneous symmetry breaking or thermalization harder to happen.

Unlike the AIII class, TRS is possible in BDI class,
while the fermion number parity symmetry $P$ is always preserved since it can always has $P^{2}=1$ by gauge fixing.
We will know that the fermion number here contains, e.g., the fermion number (or Majorana mode number) and the eigenvalue of U(1) charge.
To see this, we firstly make the transform to the Fock basis of Majorana operator,
\begin{equation} 
\begin{aligned}
\chi_{2i}=&(c_{i}+c_{i}^{\dag}),\\
\chi_{2i-1}=&-i(c_{i}-c_{i}^{\dag}),
\end{aligned}
\end{equation}
and thus 
\begin{equation} 
\begin{aligned}
c_{i}=&\frac{1}{2}(\chi_{2i}+i\chi_{2i-1}),\\
c^{\dag}_{i}=&\frac{1}{2}(\chi_{2i}-i\chi_{2i-1}).
\end{aligned}
\end{equation}
Next we introduce the fermion number operator as $\mathcal{Q}=i\chi_{2i-1}\chi_{2i}=[c_{i},c_{i}^{\dag}]=(1-2c_{i}^{\dag}c_{i})$.
In the presence of TRS,
the trivial phase of chain corresponds to Hamiltonian
\begin{equation} 
\begin{aligned}
H_{0}=\frac{-1}{2}\sum^{N}_{i}\mathcal{Q}=\frac{-1}{2}\sum^{N}_{i}i\chi_{2i-1}\chi_{2i}=\sum^{N}_{i}(c^{\dag}_{i}c_{i}-\frac{1}{2}),
\end{aligned}
\end{equation}
which is the U(1) charge operator.
While in the presence of topological Haldane order, which gives rise to topological nontrivial phase,
the Hamiltonian becomes
\begin{equation} 
\begin{aligned}
H_{1}
&=\frac{-1}{2}\sum^{N-1}_{i}i\chi_{2i}\chi_{2i+1}\\
&=\frac{-1}{2}\sum^{N-1}_{i}
(c_{i}c_{i+1}-c_{i}c_{i+1}^{\dag}+c^{\dag}_{i}c_{i+1}-c^{\dag}_{i}c_{i+1}^{\dag})\\
&=\frac{1}{2}\sum^{N-1}_{i}
(c_{i+1}c_{i}+c_{i}c_{i+1}^{\dag}+c_{i+1}c^{\dag}_{i}+c^{\dag}_{i}c_{i+1}^{\dag}).
\end{aligned}
\end{equation}
This corresponds to the a emergent topological order without the spontaneously breaking of a local order parameter.
In this phase ($H_{1}$), there is a twofold ground-state degeneracy in noninteracting case,
where the preserved TRS prevents the formation of expectation value of fermion number operator (U(1) charge)
and thus reduce the $\mathds{Z}_{8}$ symmetry to $\mathds{Z}_{2}$ (inlcuding the $\mathds{Z}_{2}$ TRS, $\mathds{Z}_{2}$ PHS, charge conjugation, etc.).
The two ground states are degenerate since the energy measurement is local,
and the order parameter becomes non-local when consider the spin,
thus spin flip symmetry between two degenerate ground states is spontaneously broken\cite{Fendley P}.
Note in BDI class with $\mathcal{T}^{2}=1$,
this trivial-to-topologically nontrivial phase transition can be avoided by strong enough local interaction,
while the weak local interaction is perturbatively irrelevant for this transition.

The fermion number parity is defined by
\begin{equation} 
\begin{aligned}
P=\prod^{N}_{i}i\chi_{2i-1}\chi_{2i},
\end{aligned}
\end{equation}
which becomes $P=\prod^{N}_{i}(-\sigma^{z}_{i})$ after Jordan Wigner transform (see Appendix.A).
By defining $N_{\chi}$ as the number of Majorana modes,
where $N_{\chi}=2N$ when $N_{\chi}$ is even,
we can rewrite the $P$ as
\begin{equation} 
\begin{aligned}
P=(i\chi_{1}c_{2})\cdot\cdot\cdot (i\chi_{N_{\chi}-1}\chi_{N_{\chi}}),
\end{aligned}
\end{equation}
for $N_{\chi}$ is even,
and
\begin{equation} 
\begin{aligned}
P=(i\chi_{1}c_{2})\cdot\cdot\cdot (i\chi_{N_{\chi}}\chi_{\infty}),
\end{aligned}
\end{equation}
for $N_{\chi}$ is odd.
An additional Majorana mode $\chi_{\infty}$ is added in the latter case
to makes it be an even-parity operator.
Thus a single Majorana mode can breaks symmetry of $P$: $P\chi P^{-1}=-\chi$, no matter its index is even or odd,
that is also why the $P$ is broken once the paired Majorana modes (in the ground state) can be gapped out.
Then following the definition of topological index $k$ in Ref.\cite{Fidkowski L},
the Hamiltonian can be written as
$H=\frac{-i}{2}\sum^{N}_{i}\chi_{2i}\chi_{2i+2k-1}$,
which restricts the real spacial distance of the quadratic coupling in the 1D chain.
Then only in the case of zero $k$, which corresponds to the trivial phase,
the ground state is completely unoccupied.
While the case of $k=1$ corresponds to the above topological Haldane order.

To see this,
we assume $N_{\chi}$ is even,
then for $k=1$,
there are two edge Majorana modes $\chi_{1}$ and $\chi_{N_{\chi}}$ in ground state and they can be
paired up to form a two-fold ground state degeneracy,
i.e., the $Z_{2}$ symmetry class.
That is because now the degenerate ground state $i\chi_{1}\chi_{N_{\chi}}$ can be gapped out 
since $\chi_{1}$ and $\chi_{N_{\chi}}$ are of different fermion parities as long as $N_{\chi}$ is even,
and that also makes them satisfy the TRS:
\begin{equation} 
\begin{aligned}
\mathcal{T}i\chi_{1}\chi_{N_{\chi}}\mathcal{T}^{-1}
=-i(-1)^{1+N_{\chi}}(-\chi_{1})(-\chi_{N_{\chi}})
=i\chi_{1}\chi_{N_{\chi}},
\end{aligned}
\end{equation}
and we found that this is valid after replacing $\mathcal{T}$ by the antiunitary one $\mathcal{T}$.
Here we note that
\begin{equation} 
\begin{aligned}
\mathcal{T}i\mathcal{T}^{-1}=-i,\\
\mathcal{T}\chi_{a}\mathcal{T}^{-1}=-(-1)^{a}\chi_{a},\\
\mathcal{K}\chi_{a}\mathcal{K}^{-1}=-(-1)^{a}\chi_{a},\\
\mathcal{T}\mathcal{K}\chi_{a}\mathcal{K}^{-1}\mathcal{T}^{-1}=\chi_{a},
\end{aligned}
\end{equation}
i.e., a single Majorana mode always satisfies the antiunitary TRS.
As the local fermionic excitation $i\chi_{1}\chi_{N_{\chi}}$ here at the edge is not being protected by TRS,
it can be fully gapped out.
In this case,
as the U(1) gauge symmetry as well as the $\mathds{Z}_{2}$ fermion number parity symmetry is broken,
the system now is of the $\mathds{Z}_{2}$ classfication:
including the $\mathds{Z}_{2}$ TRS, $\mathds{Z}_{2}$ antiunitary PHS, $\mathds{Z}_{2}$ charge conjugation symmetry.
As long as $k$ is odd, which equivalents to odd $N_{\chi}$,
there will be an anormalous component, and the corresponding topologically nontrivial order,
and the TRS interacting pair (fermionic excitation) will breaks $P$ and thus there are no more U(1) charge symmetry.
As we consider the $N_{\chi}\rightarrow\infty$ limit,
the Majorana resonance happen at zero energy (zero bias),
so resonance $i\chi_{1}\chi_{N_{\chi}}$ is a zero energy mode.
If there is no SYK interaction which couples four Majorana modes
in the same time, this hybridization of this resonance will simply decays as a exponential law with the Majorana chain size $N_{\chi}$,
since there are pairing of Majorana modes which leads to a gap in the bulk part of the Majorana chain.

While for the $k=2$ case (we still assume $N_{\chi}$ is even),
although now there are ground states occupied by $\chi_{1},\chi_{3},\chi_{N_{\chi}-2},\chi_{N_{\chi}}$,
they can be paired up into two interacting terms (two bosonic ground states) $i\chi_{1}\chi_{3}$
and $i\chi_{N_{\chi}-2}\chi_{N_{\chi}}$ without cost energies.
This is because the degeneracy between these four modes can be gapped out by interaction into two bosonic gapless edge modes,
which is allowed by the $\mathds{Z}_{4}$ TRS.
However, these two bosonic modes cannot be further gapped out,
since the two Majorana modes within each of them belong to the same fermion parity 
(all even or all odd),
and thus do not satisfy $\mathds{Z}_{2}$ TRS, e.g.,
\begin{equation} 
\begin{aligned}
\mathcal{T}i\chi_{1}\chi_{3}\mathcal{T}^{-1}
=-i(-1)^{1+3}(-\chi_{1})(-\chi_{3})
=-i\chi_{1}\chi_{3}.
\end{aligned}
\end{equation}
This corresponds to a TRS-protected local gapless edge boson mode.
Thus $P$ can only be preserved in even-$k$ case (the topologically trivial case),
where the $P$ will breaks $Z_{2}$ symmetry by a finite expectation value of U(1) conserved charge.
Similar to the SYK intercation in boundary Hamiltonian,
the topological superconductivity\cite{Jian C M,Vadimov V} and the Coulomb intercation\cite{Bi Z} 
and fluctuations in thermalized gas\cite{Chandran A}
can also gapped out the single-fermion excitations
which constituted of two Majorana operators with different fermion parity.

In the presence of even $k$,
the we can always have $P^{2}=1$ by the gauge fixing through the phase rotation.
Let the Hamiltonian being block diagonalized in each fermion parity,
\begin{equation} 
\begin{aligned}
H_{\pm}=
\begin{pmatrix}
H_{P=+1} & 0\\
0 & H_{P=-1}
\end{pmatrix},
\end{aligned}
\end{equation}
then if $N_{\chi}$ is odd, $P$ breaks, and 
we have $H_{P=+1}=H_{P=-1}$.

As we consider the replica limit $N_{\chi}\rightarrow \infty$ as required by solvable SYK model,
we use the method of matrix product state, after enforcing an entanglement bipartition to each pair of Majorana modes,
insteads of the exact diagonalization.
Note that if a perturbation (e.g., the random hopping) induces a fluctuation (extensive in $N_{\chi}$) to the saddle point
obtained by conformal perturbation theory,
the replica symmetry will be broken.
Then for even $N_{\chi}$ and odd $k$, the parity operator after bipartition 
can be reduced to a product of $k/2$ terms
\begin{equation} 
\begin{aligned}
P=(i\chi_{2}\chi_{4})\cdot\cdot\cdot (i\chi_{2k-2}\chi_{2k}).
\end{aligned}
\end{equation}
Since 
\begin{equation} 
\begin{aligned}
\mathcal{T}\chi_{2a}\mathcal{T}^{-1}=-\chi_{2a},\\
\mathcal{T}i\mathcal{T}^{-1}=-i,
\end{aligned}
\end{equation}
we have $\mathcal{T}P\mathcal{T}^{-1}=(-1)^{k/2}P$.

While for even $N_{\chi}$ and even $k$, the parity operator after bipartition 
can be reduced to a product of $(k+1)/2$ terms
\begin{equation} 
\begin{aligned}
P=(i\chi_{2}\chi_{4})\cdot\cdot\cdot (i\chi_{2k}\chi_{\infty}).
\end{aligned}
\end{equation}
By choosing
\begin{equation} 
\begin{aligned}
\mathcal{T}\chi_{\infty}\mathcal{T}^{-1}=(-1)^{(k+1)/2}\chi_{\infty},
\end{aligned}
\end{equation}
we have $\mathcal{T}P\mathcal{T}^{-1}=(-1)^{k}(-1)^{(k+1)/2}(-1)^{(k+1)/2}P=(-1)^{2k+1}P=-P$.
Thus we conclude that the $P$ satisfies TRS only when $N_{\chi}$ mod 4=0,
which is of the classes $\mathds{Z}_{4}$ or $\mathds{Z}_{8}$ in the presence of interaction (see Table.1).

\begin{table}[htbp]
	\centering
	\caption{This table shows the antiunitary TRS, TRS, level statistic, symetry classes 
and if the blocks of Hamiltonian being exchanged by the $\mathcal{T}$,
as well as the preservation of $P$ symmetry for different values of Fermion number $N$ and Majorana mode number $N_{\chi}$.
We note that in all $\mathds{Z}_{2}$ classes, $H_{P=+}=H_{P=-}$ and $H_{-\lambda_{Q}}=H_{\lambda_{Q}}$.
For $N_{\chi}$ mod 8=0 or 4, in which case $N_{\chi}$ is divisible by 4,
$\mathds{Z}_{2}$ symmetry is broken by $\mathcal{Q}$, then finite and zero $\lambda_{Q}$ correspond to $\mathds{Z}_{8}$ and $\mathds{Z}_{4}$
symmetries, respectively.
}
	\begin{tabular}{ccccccccc}
		\toprule  
$N$ mod 4 & $N_{\chi}$ mod 8 & $\mathcal{TK}^{2}$ & $\mathcal{T}^{2}$ & $\mathcal{T}H\mathcal{T}^{-1}$ & Level statistic & Class & Exchange$?$ & P symmetry\\ 
		\midrule  
0 & 0 & 1 & 1 & $H$ & GOE & $Z_{8}/Z_{4}$ & $\times$ & $\checkmark$\\
  & 1  & 1 & 1 & $H$ & GOE & $Z_{2}$ & $\checkmark$ & $\times$\\
1  & 2  & 1 &   & $H^{*}$ & GUE & $Z_{4}$ & $\checkmark$ & $\checkmark$\\
  & 3  & -1 &  -1 & $H$ & GSE & $Z_{2}$ & $\times$ & $\times$\\
2  & 4  & -1 &  -1 & $H$ & GSE & $Z_{8}/Z_{4}$ & $\times$ & $\checkmark$\\
  & 5  & -1 &  -1 & $H$ & GSE & $Z_{2}$ & $\checkmark$ & $\times$\\
3  & 6  & -1 &   & $H^{*}$ & GUE & $Z_{4}$ & $\checkmark$ & $\checkmark$\\
  & 7  & 1 &  1 & $H$ & GOE & $Z_{2}$ & $\times$ & $\times$\\
		\bottomrule  
	\end{tabular}
\end{table}

\section{Effect of strong disorder in Ising term}

As shown in Appendix.A,
the Ising Hamiltonian $H_{1}=\frac{1}{2}\sum_{i}\sigma_{i}^{y}\sigma^{y}_{i+1}$ is in an ordered phase
and has a nonlocal long-range order of $\sigma^{x}$,
As we stated above, in this case there are two degenerate ground states 
which can be gapped out by TRS-preserving interaction,
and in the mean time,
breaks $P$ symmetry (as well as the U(1) gauge symmetry).
But when $N_{\chi}$ is divisible by 4 and $N_{\chi}\rightarrow\infty$,
there will be an asymptotic degeneracy between the two lowest states,
in this case the finite expectation value of $U(1)$ gauge charge as well as the preserved $P$ symmetry 
will leads to spontaneous breaking of $\mathds{Z}_{2}$ symmetry,
including the $Z_{2}$ spin flip symmetry (thus SU(2) symmetry is also broken)
since the magnetic order is unlikely to happen in degenerated ground state.
The asymptotic degeneracy in the latter case cannot be gapped out.
To understand this,
we assume the lowest state is $|\psi_{0}\rangle$ which is even under $P$ operation,
$P|\psi_{0}\rangle=|\psi_{0}\rangle$.
Then using the definition of operator $\hat{O}$\cite{Huse D A,Fendley P}
base on the bilocalized fermion operators
\begin{equation} 
\begin{aligned}
\hat{O}=c_{1}^{\dag}+c_{1}+c_{\frac{N+1}{2}}^{\dag}-c_{\frac{N+1}{2}},
\end{aligned}
\end{equation}
we can obtain another ground state (anearby state of $|\psi_{0}\rangle$) $\hat{O}|\psi_{0}\rangle$ 
which is odd under $P$ operation:
$P\hat{O}|\psi_{0}\rangle=-\hat{O}P|\psi_{0}\rangle=-\hat{O}|\psi_{0}\rangle=-|\psi_{0}\rangle$.
Note that $\hat{O}$ is anticommute with $P$.
The energy different, which plays the role of local measurement, between these two lowest states
is  exponentially small in system size $N$.
While in trivial phase,
the fermion mode ($\prod_{j=1}^{i-1}\sigma_{j}^{z}$) creates the domain walls on the two-dimensional bonds.
In this case, $H_{1}$ contains Cooper pairings and do not have a U(1) symmetry.
In fact, $\hat{O}$ is anticommute with $P$ as long as it is fermionic,
thus $\{\hat{O},(-1)^{w}\}=0$, where $w$ is the index of domain walls,
and the topological states with $(-1)^{w}=1$ can be mapped to the one with $(-1)^{w}=-1$ through operaion of $\hat{O}$.

Then we rewrite the above $H_{1}$ which has a long-range Ising spin order in topological ordered phase,
as the disordered form: $H_{1}=\frac{1}{2}\sum_{i}J_{i}\sigma_{i}^{y}\sigma^{y}_{i+1}$,
where the variance $\overline{J_{i}^{2}}$ is rather large while $J_{i}$ is sparse and local in space.
In this case, the MBL system will generate paired spin glass order.
While the fermions (normal mode) and the generated domain walls (if exists) are being localized by the spin-glass order
$\langle \sigma_{i}^{y}\sigma^{y}_{j}\rangle$.
This local spin correlation can provides an approximately conserved quantity $\langle\sigma_{i}^{y}\rangle$
which is nonzero after long-time evolution by virtue of spin glass order,
This local conserved quantity with U(1) symmetry can prevents the thermalization.
There are two cases for the breaking of U(1) symmetry:
The first one is at the ground state critical point between the MBL paramagnet phase
(corresponds to weak disorder and high energy density) and the MBL ordered spin glass;
the second one is within the paramagnetic phase.
In both of these two cases,
the spin correlation is no a local conserved quantity,
and there will be a long-range spin-glass order (nonlocal) in first case;
While in second case,
for MBL paramagnetic phase in the bulk with zero interaction,
the domain-wall eigenstates still must be localized
(the nonlocal eigenstates decay in spin glass phase with a finite correlation length),
and the sign of nonlocal spin correlation will depends on the amount of domain walls between sites $i$ and $j$.
The long-range spin-glass order here can leads to spontaneous breaking of $\mathds{Z}_{2}$ discrete symmetries in localized states,
which is similar to bilinear terms or quadratic coupling terms like U(1) charge $\mathcal{Q}$.
In this two cases,
the thermalization is able to happen in the presence of generic (without integrability) interactions,
e.g., the SYK-type which is nonlocal and dense,
and the discrete symmetries will not be broken at thermal equilibrium.
The Ising paramagnetic phase corresponds to the above-mentioned trivial phase,
playing the role of confining potential.
Although similar in form with the fermion number parity operator,
the eigenstates in Ising paramagnetic MBL phase
does not breaks any $\mathds{Z}_{2}$ symmetries,
but they do in Ising paramagnetic extended phase\cite{Huse D A}.

For highly excited eigenstates which remain localized in MBL system,
the long-range spin-glass order is able to breaks the global $\mathds{Z}_{2}$ symmetry
(like the antiunitary PHS or other discrete symmetries),
as long as they are not being thermalized.
In conclusion, in ergodic phase,
the asymptotic degeneracy of two lowest eigenvalues,
which cannot be gapped out by SYK boundary Hamiltonian,
will breaks the $\mathds{Z}_{2}$ discrete symmetries and local order parameters,
as can be seem from the anticommutation relations between U(1) charge $\mathcal{Q}$ and the $\mathds{Z}_{2}$ discrete symmetries
stated above.
At the critical point or inside the paramagnetic phase (trivial),
the long-range correlation $\langle\sigma_{i}^{y}\sigma_{j}^{y}\rangle$
is nonconserved but still be a local order parameter,
and the correlation length $\xi$ of decaying nonlocal terms
will be related to this local order parameter through
$\langle\sigma_{i}^{y}\sigma_{j}^{y}\rangle-\langle\sigma_{i}^{y}\rangle\langle\sigma_{j}^{y}\rangle=e^{-|i-j|/\xi}$\cite{Moudgalya S}.
The topological phase can be deformed to the trivial product state only by closing the bulk gap through a
quantum critical point (not absolutely with continuous symmetry breaking),
for example, for anomalous symmetry in the case that $N_{\chi}$ is nonero mod 8,
the only way for such deformation is by closing the gap.

Note that the MBL Ising paramagnetic phase
can be further classified into two phases:
topological paramagnet and trivial paramagnet,
according to if the Ising symmetry exists or not.
The Ising symmetry reads $\prod_{j}\sigma^{z}_{j}$ which is commute with Hamiltonian in topological phase,
and it creates the domain walls,
which signals the topological nature.
When the Ising symmetry in topological phase is broken
the gapped mobility edge (with incomplete integrability in nonergodic phase)
in localized states will breaks $\mathds{Z}_{2}$ symmetries,
similar the what happen in disordered system with localized domain walls,
and tends to the thermalization by delocalization effects.
The existence of mobility edges also implies that this is an intermediate phase,
where the topological order is not being completely destroyed and thus there are local order parameter being preserved,
as stated in the following section.

\section{Local and nonlocal effects on two-site MPS}

The phenomenon that the $\mathds{Z}_{2}$ fermion parity symmetry $P$ is broken in topological nontrivial phase with odd $k$ and
TRS pair $i\chi_{1}\chi_{N_{\chi}}=i\chi_{1}\chi_{2N}$,
can only happen in boson chain,
and there exists local gapless bosonic edge modes with the symmetry-protected-topology.
The Hamiltonian $H_{1}$ turns to nonlocal Ising form after applying the Jordan-Wigner transformation,
since the order parameter in terms of spin becomes non-local.
Then we found that, not only the gapped local Hamiltonians\cite{Moudgalya S},
but also the gapped nonlocal Hamiltonian (Jordan-Wigner transformed)\cite{Fidkowski L}
can has ground states approximately described by MPS with small bond dimension.

As we consider the Majorana chain as a TRS boson spin chain after Jordan-Wigner transformation,
we assume there is a three-dimensional Hilbert space on each sites,
and the MPS tensors in each dimension can be expressed in terms of hard core boson operator (see Appendix.A)
\begin{equation} 
\begin{aligned}
A^{(1)}=\sqrt{\frac{2}{3}}\sigma^{+}=\sqrt{\frac{2}{3}}\frac{1}{2}(\sigma^{x}+i\sigma^{y})=\sqrt{\frac{2}{3}}b^{\dag},\\
A^{(2)}=-\sqrt{\frac{1}{3}}\sigma^{z}=-\sqrt{\frac{1}{3}}(2b^{\dag}b-1)=-\sqrt{\frac{1}{3}}(2c^{\dag}c-1),\\
A^{(3)}=-\sqrt{\frac{2}{3}}\sigma^{-}=-\sqrt{\frac{2}{3}}\frac{1}{2}(\sigma^{x}-i\sigma^{y})=-\sqrt{\frac{2}{3}}b.
\end{aligned}
\end{equation}
Note that here the superscripts of Pauli matries does not correspond to the spacial direction of spins,
the actual correspondence depends on the duality transformation we picked,
thus in our above basis, we can suspect that the above tensors
correspond to $S_{y}=1,0,-1$, respectively, with the bond dimension 2 and Hilbert space dimension 3.

Firstly we discuss the local case,
where we select the MPS of two neighbor sites on the bulk of the boson spin chain,
i.e., $|\psi_{A}\rangle=\sum_{m_{1},m_{2}=1,2,3}A^{(m_{1})}_{j}A^{(m_{2})}_{j+1}|m_{1},m_{2}\rangle$,
where $A_{j}=i\chi_{i}\chi_{i+2}$ and $A_{j+1}=i\chi_{i+1}\chi_{i+3}$.
Note that for $A_{j}$ and $A_{j+1}$,
each of them does not satisfy TRS and thus cannot be gapped out.
The local term $|\psi_{A}\rangle$ which follows area law entanglement and Poisson distribution
of level statistic,
can be a local eigenstate of the AKLT-type Hamiltonian\cite{Moudgalya S,Moudgalya S2}
\begin{equation} 
\begin{aligned}
H=\frac{N_{\chi}}{2}\hat{h}.
\end{aligned}
\end{equation}
This is an nonintegrable Hamiltonian 
(and thus with area-law entanglement) with period boundary condition,
where the term $\hat{h}$ in a decomposition form 
is indeed a projector from space of two spin-1 boson to that 
containing sectors with total spin angular momentum 2,1, and 0.
The level repulsion happen between levels of different symmetry classes, like the discrete symmetries and rotational SU(2) symmetry.
For these three cases, the number of degenerate states will be 5,3, and 1.
We suspect the degeneracy to be rised from the SU(N) exchange interaction instead of the geometrical frustraction
as we consider the MPS $|\psi_{A}\rangle $ as a frustration-free eigenstate,
and the magnetic order is suppressed by high degree of degeneracy.
For total spin-2 case with SU(2) symmetry,
the five degenerate states correspond to projection on $S_{y}$ axis as $m_{y}=$2, 1, 0, -1, -2,
and among them only the exchanges between $m_{y}=0$ states preserve the $\mathds{Z}_{2}$ spin flip symmetry.

The $\hat{h}$ can be chosen as translational invariant (without localization),
but is a local operator which vanishes when operating on the eigenstate $|\psi_{A}\rangle$
which is localized in energy space (with zero eigenvalue).
$\hat{h}$ can be block diagonalized into two block matrices:
first one is consisted of the target states $|\psi_{A}\rangle_{1}\in |1,m_{y}\rangle \oplus |0,0\rangle $ which are embedded into the total system,
its eigenvalues are zero in the absence of quasiparticle;
second one is consisted of the complement states $|\psi_{A}\rangle_{2}\in |2,m_{y}\rangle $.
When the $5\times 5$ Hermitian matrix spaned by states $|2,m_{y}\rangle $ is positive semi-definite (any subspace requires zero vector),
$|\psi_{A}\rangle$ are the degenerate ground states, otherwise $|\psi_{A}\rangle$ are embedded into the middle of spectrum.
This is different to the spin-1/2 chain whose ground states are consisted of dimers (singlets)
in the fermionic language,
which can also not be gapped out by the boundary interaction.
The local projector satisfies
$\hat{h}\propto \sum_{m_{y}}|2,m_{z}\rangle\langle 2,m_{y}|$
in the presence of SU(2) symmetry: including the $S_{y}$ U(1) symmetry and the spin rotational symmetry about $S_{y}$ axis.

As we mentioned above, since the two MPSs are the bosonic pairs that cannot be gapped out,
it corresponds to the even-$k$ trivial case,
which preserves the U(1) and $\mathds{Z}_{2}$ fermion parity symmetries.
That is why this can be described by pure MPS $|\psi_{A}\rangle$.
While for odd-$k$ nontrivial case, it should be described by mixed MPS $|\psi_{A}\rangle$.
In the latter case, the emergence of topological order requires the end points of the segment decoupled (unpaired),
which requires a long enough distance.

Just like the topological nontrivial phase (Haldane phase),
the degenerate ground states formed by dangling edge modes with spin can be gapped out by boundary Hamiltonian 
in nonlocal boson spin chain with open boundary condition,
the ground state of integer spin chain can be gapped away from the fermion excitations by a Haldane gap
which can stabilizes the ground state by prohibiting the transitions to excited states.
Note that this topological nontrivial phase is in contrast to the above-mentioned topological case where the topological index $N_{\chi}$ is divisible by 4
and the expectation $\langle\mathcal{Q}\rangle$ breaks $\mathds{Z}_{2}$ discrete symmetries in nonthermalized states.
While in AKLT model,
the ground states will be bounded from below into singlets (or dimers)\cite{Moudgalya S2,Shiraishi N},
which is invariant under $P$ operation.
The topological order in Haldane phase can be pretected by both the continuous rotational symmetries (like the SU(2) symmetry)
and the discrete symmetries (like the $\mathds{Z}_{2}$ symmetry),
which can be nonlocal with open boundary condition,
and follows ETH.
While the nonintegrable AKLT Hamiltonian (which is Hermitian) with period boundary condition,
contains only the diagonal real elements in the presence of U(1) symmetry,
thus the level statistic follows GOE,
but the eigenstates in AKLT system (target states) violate ETH and will not be thermalized.
This can be proved that for local observable $\hat{h}$,
we have $\langle \psi_{A}|\hat{h}|\psi_{A}\rangle=\langle \psi_{A}|(\hat{h}|\psi_{A}\rangle)=0$,
i.e., does not obey the ETH ansatz,
as long as the (locally bounded) Hilbert space dimension of each site is larger than the bond dimension.

Unlike in symmetry classes CII and AIII,
the charge conjugation symmetry ($\mathds{Z}_{2}^{\mathcal{C}'}$) is absent in AKLT chain with SU(2) symmetry,
although the U(1)$\times\mathds{Z}_{2}^{\mathcal{C}'}$ symmetry can be subgroup into SU(2).
This is reasonal base on two facts:
the spin rotational symmetry is broken in CII class,
which preserves the $\mathds{Z}_{2}$ antiunitary chiral symmetry;
the chiral symmetry is absent in AKLT chain with period boundary condition,
otherwise the chiral spin liquid,
which can be formed through uncondensated spinfull bosons,
will breaks both the TRS and fermion parity $P$ symmetry,
and leads to the topological order,
similar to that in the Haldane phase.
As we said, the existence of charge conjugation symmetry in CII and AIII classes
leads to extensive local degeneracies (by energy measurement),
and thus make the symmetry-protected topological orders be unstable against the boundary interactions.
While in AKLT chain,
to makes the edge modes delocalized (thermalized),
i.e., destroy the boundary modes in the symmetry-protected topological ordered state 
(the long-range string order in AKLT chain signals the topological nontriviality),
we can enlarge the spin rotational symmetry from SU(2) to SU(N$_{s}$) with $N_{s}>2$\cite{Hermele M},
in which case the magnetic order is being suppressed,
and the ground state degeneracy becomes extensive with system size,
which is, in terms of MPS representation,
by increasing the bond dimension,
although the MPS representation may fail for highly excited nonlocal eigenstates.
In this case,
the ground state degeneracy is not necessarily to be local,
and the extensive degeneracy signals the effect of geometrical frustration,
as long as the bond dimension is equals or larger to the single site Hilbert space dimension,
and then the $\hat{h}$ is nomore frustration-free.
But it does not means large $N_{s}$ always leads to thermalization,
as in some case large-$N_{s}$ will breaks translational invariance and thus violates the ETH,
like in spin-Peierls type ground state\cite{Harada K}.

Next we explain how the highly excited state in ergodic phase can destroy the symmetry-protected topological order
and gives rise to the pairing in Cooper channel,
and thus breaks the global U(1) symmetry.
Note that a direct result of the breaking of U(1) symmetry is the local term $\hat{h}$ becomes nondiagonal.
Firstly, as shown in Appendix.B,
there are five linearly independent (but not mutually orthogonal) excitation states with total spin 2 in the presence of SU(2).
Note that here the SU(2) symmetry only exists in the subspace spaned by the five degenerate excited states
(viewed as five basis states here),
while for the previous ground states with total spin 1,
the U(1) symmetry is obviously broken.
Then the diagonal contribution and the nondiagonal contribution,
which give rise to short range and long range order, respectively.
are,
\begin{equation} 
\begin{aligned}
&\langle 2,j,m_{y}|2,j,m_{y}\rangle\sim 1,\\
&\langle 2,j,m_{y}|2,j',m'_{y}\rangle\sim (-1)^{j+j'}\delta_{m_{y},m'_{y}}.
\end{aligned}
\end{equation}
The long range behavior in the latter case leads to the nonlocal effect,
and the excitations emerge in chain can be viewed as nonlocalized defects to the topological order,
which give rise to fluctuation and extensive entanglement entropy.
Before thermal equilibrium is arrived,
the remaining topologocal order (may not anymore protected by symmetries)
may gives rise to the intermediate effects\cite{Moudgalya S3,Chandran A,Schecter M} between MBL and ETH, incluing the mobility edge/gap.

For the above selected MPS $|\psi_{A}\rangle$, as long as the two sites are adjacent,
it is impossible for asymptotically degenerate to emerge,
thus in this case,
the transfer matrix $T=A^{*}\otimes A$ commute with the operators
$(\hat{S}\otimes{\bf 1})$ and $({\bf 1}\otimes \hat{S})$,
which correspond to excitation act on the left side and right side of the chain,
respectively,
and the quasiparticle creation operator $\hat{S}$ belongs to the subspace spaned by the five degenerate excited states as
presented in Appendix.B.
In MBL phase,
the eigenvector corresponding to the leading eigenvalue of transfer matrix is invariant under the relative operation
$(\hat{S}\otimes{\bf 1}-{\bf 1}\otimes \hat{S})|T\rangle=|T\rangle$.
This invariance vanishes for the vectors of subspace spaned by $|\psi_{A}\rangle_{1}$.

\section{$\mathds{Z}_{4}$ symmetry}

In $\mathds{Z}_{n}$ symmetry class with the corresponding generator $\hat{G}_{Z_{n}}$,
the topological order can be recognized by the Kramer degeneracy 
in the presence of TRS and other discrete symmetries,
in this case the single-site operator (local) has the following relation (in form of spectral decomposition)
\begin{equation} 
\begin{aligned}
({\bf 1}^{\otimes (i-1)}\otimes\hat{S}_{i}{\bf 1}^{\otimes (N-i)})\hat{G}_{Z_{n}}({\bf 1}^{\otimes (i-1)}\otimes\hat{S}_{i}{\bf 1}^{\otimes (N-i)})^{-1}
=e^{2i\pi/n}\hat{G}_{Z_{n}},
\end{aligned}
\end{equation}
where 
\begin{equation} 
\begin{aligned}
\hat{G}_{Z_{n}}=
\prod_{i}({\bf 1}^{\otimes (i-1)}\otimes\lambda_{i}{\bf 1}^{\otimes (N-i)}),\\
{\rm Tr}\lambda_{i}=\sum^{n-1}_{a=0}e^{2i\pi a/n}.
\end{aligned}
\end{equation}
It obvious that all the diagonal elements (eigenvalues) of $\lambda_{i}$ are local order parameters.
The operator $\hat{G}_{Z_{n}}$ commutes with the permutations $({\bf 1}^{\otimes (i-1)}\otimes\lambda_{i}{\bf 1}^{\otimes (N-i)})$,
and permutations with different $i$ commute to each other.

Next we take the $\mathds{Z}_{4}$ symmetry as an example.
In the presence of spin U(1) gauge symmetry (e.g., in CII class),
the charge reads 
\begin{equation} 
\begin{aligned}
\mathcal{Q}
=&\sum^{N}_{i}(f_{i}^{\dag}f_{i}-f_{i}f_{i}^{\dag})\\
=&\sum^{N}_{i}S^{z}_{i},
\end{aligned}
\end{equation}
with $f_{i}$ the spin-boson operator.
In terms of duality as shown in Appendix.A,
the $S_{z}$ term can be represented as
\begin{equation} 
\begin{aligned}
S_{z}=\frac{i}{2}\sum_{i}(\lambda_{i}^{\dag}-\lambda_{i}),
\end{aligned}
\end{equation}
where the measured eigenvalues in diagonal ensemble are described by
\begin{equation} 
\begin{aligned}
\lambda_{i}=\begin{pmatrix}
1 & 0 & 0 & 0\\
0 & \lambda_{1} & 0 & 0\\
0 & 0 & \lambda_{1}^{2}  & 0\\
0 & 0 & 0 & \lambda_{1}^{3} 
\end{pmatrix},
\end{aligned}
\end{equation}
where we denote $\lambda_{\alpha}=e^{2i\pi \alpha/n}$,
and thus $\lambda_{1}=e^{2i\pi/n}$ (the second largest eigenvalue).
Since $\lambda_{i}^{\dag}\equiv \lambda_{i}^{n-1}=\lambda_{i}^{3}$ for $\mathds{Z}_{4}$ class, we obtain
\begin{equation} 
\begin{aligned}
S_{z}
=&\frac{i}{2}
\begin{pmatrix}
1-1 & 0 & 0 & 0\\
0 & \lambda_{1}^{3}-\lambda_{1} & 0 & 0\\
0 & 0 & \lambda_{1}^{6}-\lambda_{1}^{2}  & 0\\
0 & 0 & 0 &  \lambda_{1}^{9}-\lambda_{1}^{3} 
\end{pmatrix}\\
=&\frac{i}{2}
\begin{pmatrix}
0 & 0 & 0 & 0\\
0 & 1 & 0 & 0\\
0 & 0 & 0  & 0\\
0 & 0 & 0 &  -1
\end{pmatrix}.
\end{aligned}
\end{equation}
We note that for large $n$,
the mirror-symmetric MBL eigenvalue can be seen,
which implies the emergence of approximately continuous non-Abelian symmetry.
In the presence of duality,
which corresponds to the half-step shift in space direction for
$\lambda_{i}$:
$\lambda_{i}\rightarrow\lambda_{i+1/2}=\sigma_{i}^{\dag}\sigma_{i+1}$,
$\lambda_{i+1/2}\rightarrow\lambda_{i+1}=\sigma_{i+1/2}^{\dag}\sigma_{i+3/2}$,
we can perform the SU(2) rotation as
\begin{equation} 
\begin{aligned}
f_{i}^{\dag}
=&\frac{1}{\sqrt{2}}(\frac{1}{\sqrt{4}}\sigma_{i}^{\dag}+\frac{1}{\sqrt{4}}i\sigma_{i+1}^{\dag})\\
=&\frac{1}{2}(\sigma_{i}^{\dag}+i\sigma_{i+1}^{\dag}),\\
f_{i}
=&\frac{1}{2}(\sigma_{i}-i\sigma_{i+1}),
\end{aligned}
\end{equation}
where we define the shift of eigenvalue in each state as
\begin{equation} 
\begin{aligned}
\sigma=
\begin{pmatrix}
0 &0 &1 & 0\\
0 &0 &0 & 1\\
0 &1 &0 & 0\\
1 &0 &0 & 0
\end{pmatrix},
\end{aligned}
\end{equation}
which satisfies $\sigma^{3}=\sigma^{\dag}=\sigma^{T}$
and $\sigma^{4}=1$.
Thus another U(1) gauge charge reads
\begin{equation} 
\begin{aligned}
\mathcal{Q}'
=&\sum^{N}_{i}(f_{i}^{\dag}f_{i}-f_{i}f_{i}^{\dag})\\
=&\sum^{N}_{i}
[\frac{1}{4}(\sigma_{i}^{\dag}\sigma_{i}-i\sigma_{i}^{\dag}\sigma_{i+1}
+i\sigma_{i+1}^{\dag}\sigma_{i}+\sigma_{i+1}^{\dag}\sigma_{i+1})
-\frac{1}{4}(\sigma_{i}\sigma_{i}^{\dag}+i\sigma_{i}\sigma_{i+1}^{\dag}
-i\sigma_{i+1}\sigma_{i}^{\dag}+\sigma_{i+1}\sigma_{i+1}^{\dag})]\\
=&\sum^{N}_{i}
[\frac{1}{4}(2-i\sigma_{i}^{\dag}\sigma_{i+1}
+i\sigma_{i+1}^{\dag}\sigma_{i})
-\frac{1}{4}(2-i\sigma_{i+1}^{\dag}\sigma_{i}
+i\sigma_{i}^{\dag}\sigma_{i+1})]\\
=&\sum^{N}_{i}
\frac{1}{2}(-i\sigma_{i}^{\dag}\sigma_{i+1}+i\sigma_{i+1}^{\dag}\sigma_{i}).
\end{aligned}
\end{equation}
In basis of $\lambda_{i}$ in $\mathds{Z}_{4}$ symmetry with discrete non-Abelian symmetry,
the spin-1 operators have a form different to the ones presented in Appendix.B
\begin{equation} 
\begin{aligned}
S^{+}
=&(3-\lambda_{1}\lambda_{i}-\lambda_{1}^{2}\lambda_{i}^{2}-\lambda_{1}^{3}\lambda_{i}^{3})\sigma^{3}\\
=&(3-e^{2i\pi/4}\lambda_{i}-e^{i\pi}\lambda_{i}^{2}-e^{3i\pi/2}\lambda_{i}^{3})\sigma^{3}\\
=&
\begin{pmatrix}
0 & 0 & 0 & 3-\lambda_{1}-\lambda_{1}^{2}-\lambda_{1}^{3}\\
0 & 0 & 3-\lambda_{2}-\lambda_{1}^{4}-\lambda_{1}^{6} & 0\\
3-\lambda_{3}-\lambda_{1}^{6}-\lambda_{1}^{9} & 0 & 0  & 0\\
0 & 3-\lambda_{4}-\lambda_{1}^{8}-\lambda_{1}^{12} & 0 &  0
\end{pmatrix}\\
=&
\begin{pmatrix}
0 & 0 & 0 & 1\\
0 & 0 & 1 & 0\\
1 & 0 & 0  & 0\\
0 & 0 & 0 &  0
\end{pmatrix},\\
S^{-}=&(S^{+})^{\dag}\\
=&
\begin{pmatrix}
0 & 0 & 1 & 0\\
0 & 0 & 0 & 0\\
0 & 1 & 0  & 0\\
1 & 0 & 0 &  0
\end{pmatrix}.
\end{aligned}
\end{equation}
Note that $\lambda_{1}^{2}=\lambda_{1}^{6}$, $\lambda_{1}=\lambda_{1}^{9}$.

As we stated above, except in the CII class,
the antiunitary TRS (independent of the fermion number parity as shown in Eq.(18)) 
can be preserved until a Zeeman term is being added to the system through the magnetic field,
since unitary $\mathcal{T}$ acts directly on the spin degree of freedom.
In symmetries-protected topological order,
the seperable degenerate ground state provides the Kramers pairs with TRS,
and the global U(1) symmetry can be broken,
in charge channel,
by the well-defined excitations above a bulk gap,
or in spin channel,
by the formation of spin singlet superconductivity (s-wave).

In above, $\lambda_{i}$ provides the measurement of eigenvalues on each state,
which provides the exact eigenvalues only in the asymptotic degenerate limit,
i.e., long-time limit $t\rightarrow\infty$ in unitary evolution of strong nonlocal interaction,
which is obviously not the case for $\mathds{Z}_{4}$ symmetry,
that is why the shift $\sigma$ is necessary.
In MBL case the symmetry-protected topological order has a unique leading eiegnvalue $\lambda_{0}>1$.
in terms of parafermions with $\mathds{Z}_{4}$ generalization:
$\psi_{2j-1}=(\prod^{j-1}_{i=1}\lambda_{i})\sigma_{j}=\lambda_{1}^{\sum^{j-1}_{i}S^{z}}\sigma_{j}$,
$\psi_{2j}=\lambda_{1}^{3/2}\psi_{2j-1}\sigma_{j}$,
we have $\psi_{2j-1}\psi_{2j}=\lambda_{1}\psi_{2j}\psi_{2j-1}$,
which is because $(\prod^{j-1}_{i=1}\lambda_{i})\sigma_{j}\lambda_{j}=\lambda_{1}\lambda_{j}(\prod^{j-1}_{i=1}\lambda_{i})\sigma_{j}$,
thus we can further obtain
 $\psi_{2j-1}\lambda_{j}=\lambda_{1}\lambda_{j}\psi_{2j-1}$,
 $\lambda_{j}=(\lambda_{1})^{-3/2}\psi_{2j-1}^{\dag}\psi_{2j}$.
Although the U(1) symmetry is broken for a Hamiltonian containing both $\lambda_{i}$ and $\sigma_{i}$ terms,
the related discrete symmetries would be broken for short range parafermionic operation\cite{O Brien E,Fendley P}.
Once the shift to eigenvalues of transfer matrix leads to symmetry-protected degeneracy,
the nonlocal term turns to local one with infinite correlation length\cite{Orus R},
similar to what happen in Wishart SYK model
with extensive residual entropy\cite{Iyoda E}.

The eigenvalue of U(1) charge $Q=\frac{i}{2}\sum^{N}_{i}(\lambda_{i}^{\dag}-\lambda_{i})$ is integer only when $N$ even,
its dual $\hat{Q}=\frac{i}{2}\sum^{N}_{i}(\sigma_{i+1}^{\dag}\sigma_{i}-\sigma_{i}^{\dag}\sigma_{i+1}\lambda_{i})$
can be decomposed into three parts\cite{Vernier E}:
$\hat{Q}=Q_{0}+Q^{+}+Q^{-}$.
We define $\langle Q\rangle=Q'/Q$ as the ratio between eigenvalues of $Q$ before and after operated by $\hat{Q}$,
then $\langle Q_{0}\rangle=1$ as it preserves the U(1) symmetry.
It is easy to verify $[Q,\hat{Q}]=4(Q^{+}-Q^{-})=[Q,Q^{+}]-[Q,Q^{-}]$,
with $Q^{+}=\frac{i}{2}\sum_{i}\sigma_{i+1}^{\dag}\sigma_{i},\ Q^{-}=\frac{i}{2}\sum_{i}\sigma_{i}^{\dag}\sigma_{i+1}$.
Since $\sigma_{j}\lambda_{k}=[\delta_{jk}\lambda_{1}+(1-\delta_{jk})]\lambda_{k}\sigma_{j}$ ($\lambda_{k}=\prod^{k-1}_{i}\lambda_{i}$),
$[Q,Q^{\pm}]=\pm 4Q^{\pm}$,
the eigenvalues of $Q^{+},Q^{-},c_{i}^{\dag},c_{i}$
are $(1+\frac{4}{\langle Q\rangle})^{\alpha},(1-\frac{4}{\langle Q\rangle})^{\alpha},
(1+\frac{1}{\langle Q\rangle})^{\alpha},(1-\frac{1}{\langle Q\rangle})^{\alpha}$, respectively,
with $\alpha=0,1,2,3$,
with $\alpha=0$ corresponds to the unique leading eigenvalue
in the symmetry-unbroken phase.

\section{Unitary time evolution in ergodic phase}

The two-point correlation can be generated through unitary long time evolution of
the uncorrelated initial product states, and the dynamics depends on the energy density of initial product state.
The long range character (nonlocal) appears in symmetry-broken system,
which allows the continuous deformation from the symmetry-protected topological phase (described by the matrix product state)
to the trivial product state.
In the topological phase with short-range spin order (like the Ising system),
the U(1) gauge symmetry is broken while the $\mathds{Z}_{2}$ discrete symmetries are preserved
such that the unpaired dangling Majorana edge zero modes can be localized in the edge of the gapped systems
(gapped out by boundary interaction terms).
In MBL spin glass order phase with strong disorder,
the nonlocal term, like the single domain wall,
will be localized with a finite localization length.
The resulting local conserved quantity can be viewed as a isolated quantum state with nonergodicity.
While in the ground state critical or the extended paramagnet phase,
the normal fermion mode will no more be a local conserved quantity,
and the spin glass order becomes long-ranged at critical point which can
 breaks the $\mathds{Z}_{2}$ symmetry in highly excited localized eigenstates\cite{Huse D A}.

Next we discuss the unitary time evolution in which case the TRS is broken 
(due to the symmetry under cyclic permutation) similar to the case in AIII class.
In nonlocal case,
the above local term $\hat{h}_{j,j+1}$ can be reformed into a nonlocal level decomposition $\hat{h}_{s_{j},s_{j+1}}$
which is related to the spectral form factor through $K(t)=({\rm Tr}\hat{h}_{s_{j},s_{j+1}})^{2}$
where $t$ stands for unitary time evolution.
Firstly, since the local Hilbert space dimension is $d=3$
for a spin-1 boson chain as we stated above,
the local term $\hat{h}_{j,j+1}$ is a $d^{2}\times d^{2}=9\times 9$ matrix,
while the nonlocal term $\hat{h}_{s_{j},s_{j+1}}$ is a $d^{N}\times d^{N}=3^{N}\times 3^{N}$ matrix ($N=N_{\chi}/2$,
and $d^{N}=3^{N}$ is the Fock space dimension or the spectrum level number here),
they are related by $\hat{h}_{s_{j},s_{j+1}}=\hat{h}_{j,j+1}^{\otimes N/2}$ for period boundary condition with $N$ even.
The translational symmetry is preserved here.
Thus $\hat{h}_{s_{j},s_{j+1}}$ is a Haar-random unitary matrix, with the Haar measure over $O(3^{N})$.
$\hat{h}_{s_{j},s_{j+1}}\hat{h}^{*}_{s_{j},s_{j+1}}={\bf 1}$,
and its distribution satisfies GUE in the thermalized state
when $3^{N}$ is large enough,
in which case the elements of unitary matrix can be regarded as independent Gaussian variables,
just like the boundary SYK interactions.
This can also be verified through the QR decomposition of a standard Gaussian distributed matrix\cite{Garratt S J,Garratt S J2}
or the polar decomposition of the unitary one\cite{Brouwer P W}.
Base on this, the spectral form factor reads
\begin{equation} 
\begin{aligned}
K(t)=&[\sum^{t-1}_{s_{j},s_{j+1}=0}\prod^{t-1}_{j=0}\hat{h}_{s_{j},s_{j+1}}][\sum^{t-1}_{s_{j},s_{j+1}=0}\prod^{t-1}_{j=0}\hat{h}^{*}_{s_{j},s_{j+1}}]\\
=&\sum^{t-1}_{s_{j},s_{j+1}=0}\prod^{t-1}_{j=0}[\delta_{s_{j},s_{j+1}}+(1-\delta_{s_{j},s_{j+1}})e^{i(\varepsilon_{j}-\varepsilon_{j+1})t}],
\end{aligned}
\end{equation}
where $s_{j}=s_{j+1}$ and $s_{j}\neq s_{j+1}$ correspond to formation of pairs and domain walls, respectively.
and this is related to the Fourier transform of two-point correlation.
$\hat{h}_{s_{j},s_{j+1}}$ can be rewitten in the level decomposition form as $\sum_{n}e^{i E_{n}t}|n\rangle\langle n|$ after introducing the unitary time evolution,
as shown in Fig.1.

For a closed path of unitary time evolution,
we have $s_{0}=s_{t}$.
Different to the Poisson statistics,
the form factor $\langle K(t)\rangle \sim t$ with $0\le t \le 3^{N}$ once it enters the ergodic phase
where the eigenvalues of $K(t)$ asymptotically degenerate in long time limit,
and then decays gradually to Heisenberg time $\langle K(t)\rangle \sim 3^{N}$ at $t\ge 3^{N}$,
which follows the Wigner-Dyson statistics.
In the latter case,
the level spacing is defined by $\Delta=\frac{2\pi}{3^{N}}$.
Since the above expression of $K(t)$ can be rewritten in terms of average as
\begin{equation} 
\begin{aligned}
K(t)=&[t^{2}\overline{(\prod^{t-1}_{j=0}\hat{h}_{s_{j},s_{j+1}})}]^{2},
\end{aligned}
\end{equation}
thus the average $\overline{(\prod^{t-1}_{j=0}\hat{h}_{s_{j},s_{j+1}})}\sim 1/t^{3/2}$ in the range $1\le t \le 3^{N}$,
and $\overline{(\prod^{t-1}_{j=0}\hat{h}_{s_{j},s_{j+1}})}\sim 3^{N/2}/t^{2}$ (which is 
vanishingly small in thermadynamic limit or long time limit) in the range $t \ge 3^{N}$.
For $\overline{(\prod^{t-1}_{j=0}\hat{h}_{s_{j},s_{j+1}})}\sim 0$ in the thermadynamic limit,
since the distributions of each term of $\hat{h}_{s_{j},s_{j+1}}$ are independent and identically
(follows the same kind of distribution),
we have $\overline{(\prod^{t-1}_{j=0}\hat{h}_{s_{j},s_{j+1}})}=\overline{\hat{h}_{s_{0},s_{1}}}\ \overline{\hat{h}_{s_{2},s_{3}}} \cdot\cdot\cdot\sim 0$,
thus the entries of Haar-random distributed O(N) matrix $K(t)$ can be approximately treated as Gaussian variables with zero mean in this case.
In terms of the single site Gaussian distribution
\begin{equation} 
\begin{aligned}
K(t\ge 3^{N})
=&[\sum_{i}\sqrt{\frac{\sqrt{2\pi}}{3^{N}\sigma}e^{-\frac{(\varepsilon_{i}-\varepsilon_{0})^{2}}{2\sigma^{2}}}}e^{3^{N}it\varepsilon_{i}}]^{2}\\
=&\frac{\sqrt{2\pi}}{3^{N}\sigma}[\sum_{ij=1}^{3^{N}}
e^{-\frac{\varepsilon_{i}^{2}}{4\sigma^{2}} } 
e^{-\frac{\varepsilon_{j}^{2}}{4\sigma^{2}} }    e^{3^{N}it(\varepsilon_{i}-\varepsilon_{j})}]^{2}\\
\sim 3^{N},
\end{aligned}
\end{equation}
where we assume zero mean ($\varepsilon_{0}=0$), and the Gaussian width reads $\frac{\sqrt{2\pi}}{3^{N}\sigma}=3^{-N}$,
i.e., the inverse level number.
Here $\sqrt{2\pi}\sigma$ denotes the energy range (many body spectrum width) and also the disorder average (or deviation of random walk)
The random matrix theory is satisfied for narrow Gaussian width $\Delta\ll \sqrt{2\pi}\sigma\ll 1$.

In terms of the Fourier-transformed two-point correlation 
we have
\begin{equation} 
\begin{aligned}
\langle K(t)\rangle
=&\sum_{ij=1}^{3^{N}}e^{it(\varepsilon_{i}-\varepsilon_{j})} \\
=&3^{2N}\overline{e^{it(\varepsilon_{i}-\varepsilon_{j})} }\\
=&3^{2N}\int d\varepsilon_{1}\cdot\cdot\cdot d\varepsilon_{3^{N}}P(\varepsilon_{1}\cdot\cdot\cdot \varepsilon_{3^{N}})
e^{it(\varepsilon_{i}-\varepsilon_{j})} \\
=&3^{N}(1+{\rm cos}(\varepsilon t)),
\end{aligned}
\end{equation}
where a normalization constant $3^{-N}$ is contained in the distribution function.
Note that here the overline is the average over the levels.
For case that the exponential term $e^{it(\varepsilon_{i}-\varepsilon_{j})}$ follows Gaussian distribution,
we have
\begin{equation} 
\begin{aligned}
P(\varepsilon_{1}\cdot\cdot\cdot \varepsilon_{3^{N}})=
3^{-N}e^{-\frac{\varepsilon_{i}^{2}}{4\sigma^{2}} } 
e^{-\frac{\varepsilon_{j}^{2}}{4\sigma^{2}} }  ,
\end{aligned}
\end{equation}
with $3^{-N}$ plays the role of normalization constant which make $\int d\varepsilon_{1}\cdot\cdot\cdot d\varepsilon_{3^{N}}
P(\varepsilon_{1}\cdot\cdot\cdot \varepsilon_{3^{N}})=1$.
Note that $\sum_{ij}e^{-\frac{\varepsilon_{i}^{2}}{4\sigma^{2}} } 
e^{-\frac{\varepsilon_{j}^{2}}{4\sigma^{2}} } =(1+2\sum^{\infty}_{m=1}e^{-3^{N}m^{2}/4})^{2}\sim 3^{N}$
with the spectrum width $\sigma=3^{-N/2}$ (which is consistent with the prefactor of SYK$_{2}$ model\cite{wusyk}).
This is the case discussed in above paragraph,
and in fact,
for large enough $3^{N}$,
the local terms (observables) in each level are independent random variables,
no matter what kind of average is applied,
which satisfies the ETH.

If we further average over the sites of each level,
the above average (over levels) term $(1+{\rm cos}(\varepsilon t))$ should be rewritten as
$\prod^{N}_{n=1}(1+{\rm cos}(\varepsilon_{n} t))$ (here $n$ denotes site index, and $N\equiv N_{\chi}/2$).
Unlike the above average process over the levels where
there is not product between different independent variables (i.e., the quasienergies $\varepsilon_{i}$ and $\varepsilon_{j}$),
the product between different independent variables appears during the average over sites,
thus the distribution function takes the form (we restrict to the two body correlation)
\begin{equation} 
\begin{aligned}
P(\varepsilon_{1}\cdot\cdot\cdot \varepsilon_{N})=
C\prod_{n}^{N}e^{-\frac{\varepsilon_{n}^{2}}{2\sigma^{2}} } 
\prod_{n<m}|\varepsilon_{n}-\varepsilon_{m}|^{2},
\end{aligned}
\end{equation}
where $\sigma=N^{-1/2}$,
and the spectral form factor now reads
\begin{equation} 
\begin{aligned}
\langle K(t)\rangle
=3^{N}\int^{\sigma/2}_{-\sigma/2} d\varepsilon_{1}\cdot\cdot\cdot d\varepsilon_{N}
P(\varepsilon_{1}\cdot\cdot\cdot \varepsilon_{N})
\prod^{N}_{n=1}(1+{\rm cos}(\varepsilon_{n} t)).
\end{aligned}
\end{equation}
Applying the disorder average,
we have
\begin{equation} 
\begin{aligned}
\overline{\prod_{n<m}|\varepsilon_{n}-\varepsilon_{m}|^{2}}
=\overline{|\varepsilon_{n}-\varepsilon_{m}|^{2}}\ \overline{|\varepsilon_{n'}-\varepsilon_{m'}|^{2}}\cdot\cdot\cdot
=(\frac{\delta\varepsilon^{2}}{4N})^{N^{2}/2},
\end{aligned}
\end{equation}
where $\delta\varepsilon$ is constant quasienergy separation
and we assume the separations $|\varepsilon_{n}-\varepsilon_{m}|$ are independent with each other.

As now we obtain the variance $\overline{|\varepsilon_{n}-\varepsilon_{m}|^{2}}=\frac{\delta\varepsilon^{2}}{4N}$,
i.e., the separation no more depends on the site index,
we can view the expression of probability density function $P(\varepsilon_{1}\cdot\cdot\cdot \varepsilon_{N})$
as the mean value $\overline{\prod_{n<m}|\varepsilon_{n}-\varepsilon_{m}|^{2}}$,
i.e., 
the distribution of $\prod_{n<m}|\varepsilon_{n}-\varepsilon_{m}|^{2}$ is just $\prod_{n}^{N}e^{-\frac{\varepsilon_{n}^{2}}{2\sigma^{2}} } $.
So through the probability constaint
\begin{equation} 
\begin{aligned}
\int d\varepsilon_{1}\cdot\cdot\cdot d\varepsilon_{N} P(\varepsilon_{1}\cdot\cdot\cdot \varepsilon_{N})=
\int d\varepsilon_{1}\cdot\cdot\cdot d\varepsilon_{N} 
C\prod_{n}^{N}e^{-\frac{\varepsilon_{n}^{2}}{2\sigma^{2}} } 
\prod_{n<m}|\varepsilon_{n}-\varepsilon_{m}|^{2}\\
=C\overline{\prod_{n<m}|\varepsilon_{n}-\varepsilon_{m}|^{2}}=1,
\end{aligned}
\end{equation}
we can obtain the normalization value as $C=(\frac{\delta\varepsilon^{2}}{4N})^{-N^{2}/2}$.

At large time $t\ge 3^{N}$ in thermadynamical limit with large $N$,
the spectral form factor enters the plateau region,
and there are level repulsion and asymptotic degeneracy (between two lowest eigenstates) in ergodic phase.

In conclusion,
the $\mathds{Z}_{n}$ symmetry is similar to the time periodicity in Floquet system,
and the duality in terms of unitary time evolution is in fact the half-step transition of interger time, $t\rightarrow (t+1/2)$,
which can be achieved by the $9\times 9$ unitary matrix $\hat{h}_{j,j+1}$ for $j$ even,
or $S\hat{h}_{j,j+1}S^{-1}$ for $j$ odd where $S$ is the half-step time transition operator\cite{Garratt S J2}.
The noncommuting character between $S$ and $\hat{h}_{j,j+1}$ in the latter case is similar to that between the shift
$\sigma$ and eigenvalue $\lambda$ as stated above.
Although the Floquet operator $\prod^{t-1}_{j=0}\hat{h}_{s_{j},s_{j+1}}$ is Haar-random,
its trace over single-site orbits is Gaussian distributed in large-$3^{N}$ limit,
and thus the spectral form factor which is the product of Floquet operator and its complex conjugation
no more follows Gaussian distribution.
So the modified nonlocal AKLT Hamiltonian in ergodic phase
reads $H=\sum^{M}_{n,m}\hat{h}_{m,n}=\sum^{M}_{n,m}z_{m,n}|m\rangle\langle n|$
with the large degenarate number $M$,
and a signal of the quantum chaos is that the Gaussian distrbuted $z_{m,n}$ satisfies 
$\overline{z_{m,n}^{2}}=\overline{z_{n,m}^{2}}=\frac{1}{2}\overline{z_{n,n}^{2}}\ (m\neq n)$,
i.e., the variance of off-diagonal elements of a Hamiltonian satisfies random matrix theory is half of the variance of diagonal elements.
Besides, the symmetry-protected topological state can be transformed to trivial product state
only through eigenstate phase transition in the absence of symmetry breaking,
while such eigenstate-eigenstate fluctuation introduced by eigenstate phase transition is suppressed
by the large degenerate $M$.
Here $M$ is related to bond dimension by $M^{-1}=\sum^{3^{N}}_{\alpha=0}\frac{1}{2^{3^{N}}M}\begin{pmatrix}3^{N}\\
\alpha\end{pmatrix}$, with single-mode bond dimension 2 and single-site Hilbert space dimension 3.
The extensive entanglement entropy can be obtained by large number of Fock space dimension $3^{N}$\cite{Moudgalya S3}
or by introducing the spacial fluctuation at low energy density\cite{Chandran A}.

\section{Comparation between SYK, AKLT, and the biquadratic (Wishart) SYK models}

Similar to the $0+1$D boundary SYK modes in 1D Majorana fermion chain,
or the one generalized to $1+1$D in 1D chiral Majarana fermion chain\cite{Lian B},
the AKLT model is nonintegrable although it preserves both the continuous and discrete symmetries,
while the Wishart SYK model as generalized from Richardson-Gaudin model is integrable\cite{Iyoda E,Pehlivan Y}.
The most important difference is that in AKLT model,
the spin "charge" $S^{z}$ can be a conserved quantity in the diagonal ensemble with the coefficient with local term $\hat{h}_{m,n}$
as $z_{m,n}=z\delta_{m,n}$,
while in Wishart SYK model, the $S^{z}$ is not a conserved quantity,
i.e., it is not independent with the constructed SU(M) Hamiltonian
\begin{equation} 
\begin{aligned}
H=&-\sum_{ij}^{N}C_{i}C_{j}{\bf S}_{i}\cdot{\bf S}_{j}\\
=&-(\sum_{i}^{N}C_{i}S^{+}_{i})(\sum_{j}^{N}C_{j}S^{-}_{j}),
\end{aligned}
\end{equation}
where $S^{+}$ and $S^{-}$ are the pairing spin-$S$ operators ($M=2S+1$) which can be constructed 
in the basis of the the dual charge of $Q$\cite{Iyoda E,Pehlivan Y}
and they satisfy $[S^{+}_{i},S_{j}^{-}]=\delta_{ij}S^{z}_{i}, [S^{z}_{i},S^{\pm}_{j}]=\pm \delta_{ij}S^{\pm}_{i}$
(after duality transformation we will see that the $H$ is indeed a biquadratic model which can be obtained by deforming the AKLT model\cite{Moudgalya S}).
Then in the case that $C_{i}$ and $C_{j}$ are both normally distributed variables 
($\overline{C_{i}^{2}}=1/N$) and independent with each other,
the integrability can be revealed by the conserved (with respect to $H$) operator $\hat{P}_{i}$
whose detail expression is presented in Refs.\cite{Iyoda E,Pehlivan Y},
and thus the integrable system can be described by $H=\sum^{N}_{i}C_{i}^{2}\hat{P}_{i}$ with $[\hat{P}_{i},H]=0$.
So the integrability of Wishart SYK model originates from the conserved $\hat{P}_{i}$ and unconserved $S^{z}$
in the new basis.
For Wishart SYK model, a large number of degeneracy requires large $N$ which leads to unextensive entanglement entropy,
while an extensive entanglement entropy requires large bond dimension.

So we conclude that,
for nonintegrable AKLT model,
we have $[H,\hat{h}_{i}]\neq 0$ and $[H,\sum_{i}\hat{h}_{i}]=0$
which means the number of eigenstates of $\hat{h}_{i}$ does not affected by the total eigenvalue of $H$;
while for integrable Wishart model in new basis,
we have $[H,\hat{h}_{i}]= 0$ (the projector $\hat{h}_{i}$ is a symmetry of the system) and $[H,\sum_{i}\hat{h}_{i}]\neq 0$
which means as the eigenvalue of $H$ increases,
the number of eigenstates of $\hat{h}_{i}$ increase, and the eigenvalues in each eigenstate is invariant.
Both of these two cases violate the ETH,
and for AKLT model,
 $\hat{h}_{i}$ can have zero expectation value when acting on state $|\psi_{A}\rangle$: $\hat{h}_{i}|\psi_{A}\rangle=0$,
as long as the dimension of subspace spanned by $|\psi_{A}\rangle $ is smaller than that of Hilbert space
as that this subspace can be bounded (localized in spectrum) by the Hilbert space.
While for Wishart model,
since $H$ is semi-positive definite,
if we have $\hat{h}_{i}|\psi_{A}\rangle\neq 0$,
then $|\psi_{A}\rangle $ must be an excited state insteads of ground state. 

We further note that,
the chirality will gives rise to the edge zero mode\cite{Fendley P} or edge excitations\cite{Moudgalya S,Hu Y} in interacting systems with $\mathds{Z}_{n}$ symmetry.
For example, in AKLT model with SU(2) symmetry,
the $\mathds{Z}_{2}$ symmetry (parafermions) is generated by the coset
which is of the form $\frac{|1,m_{y}\rangle\otimes |1,m_{y}\rangle}{|1,m_{y}\rangle\oplus |1,m_{y}\rangle}
=\frac{|2,m_{y}\rangle\oplus |1,m_{y}\rangle\oplus |0,m_{y}\rangle}{|1,m_{y}\rangle\oplus |1,m_{y}\rangle}$
(see Sec.6 and Appendix.B),
i.e., the ratio between the enlarged Hilbert space with the previous one.
In the presence of large degeneracy with $SU(M\rightarrow\infty)$ (which corresponds to $\mathds{Z}_{M}$ asymptotic degeneracy)
and dense interactions, i.e., 
the numerator of the above coset becomes $\otimes^{N}|1,m_{y}\rangle$ with the fermion number $N\rightarrow\infty$,
the system will exhibits many-body quantum chaos.

So the reason why the Wishart SYK model can be integrable even in large $M$ limit
is that it has small number (only two) of coupled states
(each of which can be expressed by indices-independent terms, like U(1) charge $Q$);
While for a standard SYK$_{n}$ model,
there are $N^{n}$ such coupled states,
e.g., for SYK$_{4}$ with $i,j,k,l=1\cdot\cdot\cdot N$ there are $N^{4}$ such states,
which suppress the fluctuation of the randomly distributed coupling,
and in fact the boson flavor $M$ also contributes to this stabilization effect,
and in the mean time, enlarge the Hilbert space.

\section{Appendix.A:  Jordan Wigner transformation
}

According to the definition of Hermitian Majorana operator,
\begin{equation} 
\begin{aligned}
&\chi_{2i}=c_{i}+c_{i}^{\dag},\\
&\chi_{2i-1}=-i(c_{i}-c_{i}^{\dag}),
\end{aligned}
\end{equation}
and hard core boson
\begin{equation} 
\begin{aligned}
&b_{i}^{\dag}=\frac{1}{2}(\sigma^{x}_{i}+i\sigma^{y}_{i}),\\
&b_{i}=\frac{1}{2}(\sigma^{x}_{i}-i\sigma^{y}_{i}),
\end{aligned}
\end{equation}
we can write the Pauli matrices as
\begin{equation} 
\begin{aligned}
\sigma^{x}_{i}=b_{i}+b_{i}^{\dag},\\
\sigma^{y}_{i}=i(b_{i}-b_{i}^{\dag}),\\
\sigma^{z}_{i}=2b_{i}^{\dag}b_{i}-1.
\end{aligned}
\end{equation}
Then the fermion creation and annihilation operator can be related to hard core boson through
\begin{equation} 
\begin{aligned}
c_{i}=e^{i\phi_{i}}b_{i},\\
c_{i}^{\dag}=b^{\dag}e^{-i\phi_{i}},\\
b_{i}=e^{-i\phi_{i}}c_{i},\\
b_{i}^{\dag}=c^{\dag}e^{i\phi_{i}},\\
\end{aligned}
\end{equation}
where (using the relation $(AB)^{-1}=B^{-1}A^{-1}$)
\begin{equation} 
\begin{aligned}
e^{i\phi_{i}}=e^{i\pi\sum^{i-1}_{j=1}b_{j}^{\dag}b_{j}}=\prod^{i-1}_{j=1}e^{i\pi b_{j}^{\dag}b_{j}}=\prod^{i-1}_{j=1}(-\sigma_{j}^{z}),\\
e^{-i\phi_{i}}=(\prod^{i-1}_{j=1}(-\sigma_{j}^{z}))^{-1}=\prod^{i-1}_{j=1}(-\sigma_{j}^{z}).
\end{aligned}
\end{equation}
Thus we have
\begin{equation} 
\begin{aligned}
{\rm det} c_{i}=(-1)^{i-1} {\rm det} b_{i}=(-1)^{\sum_{j}c_{j}^{\dag}c_{j}} {\rm det} b_{i}.
\end{aligned}
\end{equation}

Using the anticommutation and commutation relations for $b$ and $c$, respectively, we have
\begin{equation} 
\begin{aligned}
b_{i}b_{i+1}=c_{i}c_{i+1}=b_{i+1}b_{i}=-c_{i+1}c_{i},\\
b_{i}^{\dag}b_{i+1}=-c_{i}^{\dag}c_{i+1}=b_{i+1}b_{i}^{\dag}=c_{i+1}c_{i}^{\dag},\\
b_{i+1}^{\dag}b_{i}=-c_{i+1}^{\dag}c_{i}=b_{i}b_{i+1}^{\dag}=c_{i}c_{i+1}^{\dag},\\
b_{i+1}^{\dag}b_{i}^{\dag}=c_{i+1}^{\dag}c_{i}^{\dag}=b_{i}^{\dag}b_{i+1}^{\dag}=-c_{i}^{\dag}c_{i+1}^{\dag},\\
b_{i}^{\dag}b_{i}=c_{i}^{\dag}c_{i}=b_{i}b_{i}^{\dag}=-c_{i}c_{i}^{\dag},
\end{aligned}
\end{equation}
we can obtain
\begin{equation} 
\begin{aligned}
\sigma_{i}^{y}\sigma_{i+1}^{y}
&=i(b_{i}-b_{i}^{\dag})i(b_{i+1}-b_{i+1}^{\dag})\\
&=-(b_{i}b_{i+1}-b_{i}b_{i+1}^{\dag}-b_{i}^{\dag}b_{i+1}+b_{i}^{\dag}b_{i+1}^{\dag})\\
&=-(c_{i}c_{i+1}+c_{i+1}^{\dag}c_{i}+c_{i}^{\dag}c_{i+1}+c_{i+1}^{\dag}c_{i}^{\dag})\\
&=(c_{i+1}c_{i}+c_{i}c_{i+1}^{\dag}+c_{i+1}c_{i}^{\dag}+c_{i}^{\dag}c_{i+1}^{\dag}),
\end{aligned}
\end{equation}
thus it is easy to verify that
\begin{equation} 
\begin{aligned}
H_{1}
=&\frac{-i}{2}\chi_{2i}\chi_{2i+1}\\
=&\frac{1}{2}(c_{i+1}c_{i}+c_{i}c_{i+1}^{\dag}+c_{i+1}c_{i}^{\dag}+c_{i}^{\dag}c_{i+1}^{\dag})\\
=&\frac{1}{2}\sigma_{i}^{y}\sigma_{i+1}^{y},
\end{aligned}
\end{equation}
i.e., the Ising Hamiltonian.
While the trivial phase is
\begin{equation} 
\begin{aligned}
H_{0}
=&\frac{-i}{2}\chi_{2i-1}\chi_{2i}\\
=&\frac{-1}{2}(1-2c_{i}^{\dag}c_{i})\\
=&\frac{1}{2}\sum_{i}\sigma_{i}^{z},
\end{aligned}
\end{equation}
where $\sigma^{z}_{i}=2b_{i}^{\dag}b_{i}-1=2c_{i}^{\dag}c_{i}-1$.

Also, we can prove that 
$H=-\sum^{(N_{\chi}-2)/2}_{i=1}i\chi_{2i-1}\chi_{2i+2}$
as mentioned in Sec.1,
which starts the long-range pairing from the first Majorana fermion,
can also be transformed to Ising form
$H=-\sum^{(N_{\chi}-2)/2}_{i=1}\sigma^{x}_{i}\sigma^{x}_{i+1}$.
In this case, the pairing that can be gapped out is $i\chi_{2}\chi_{N_{\chi}-1}=i\chi_{2}\chi_{2N-1}$.
The Ising spin order signals the topological order,
and since U(1) symmetry is broken in Ising chain with $N_{\chi}\rightarrow \infty$,
the system preserves the only the $\mathds{Z}_{2}$ symmetry until $N_{\chi}$ is divisible by 4.
The topological order of Majonara chain can be revealed by the existence of
two exponentially (in system size)-decayed-to-zero edge modes,
which are unpaired as their hybridization also decays exponentially.
Further, the topological order requires the bulk part of Majorana chain to be gapped
(e.g., the degenerate ground states with different parity)
so that the zero mode can be localized in edge.

\section{Appendix.B}

The parent Hamiltonian of localized eigenstate described by a MPS of two neighbor site on boson spin chain
can be written in terms of projectors
which projecting the Hilbert space of two spin-1 bosons (without the singlet)
into the space with total spin angular momentum 2,1, and 0.
This process can be written in terms of states $|S,m_{y}\rangle$
($S$ is the total spin angular momentum and $m_{y}$ is its projection on $y$ aixs)
\begin{equation} 
\begin{aligned}
\begin{pmatrix}
|1,1\rangle \\
|1, 0\rangle \\
|1,-1\rangle
\end{pmatrix}\otimes
\begin{pmatrix}
|1,1\rangle \\
|1, 0\rangle \\
|1,-1\rangle
\end{pmatrix}
=|2,\pm 2\rangle\oplus |2,\pm 1\rangle \oplus  |2,0\rangle
\oplus |1,\pm 1\rangle\oplus |1,\pm 0\rangle \oplus |0,0\rangle.
\end{aligned}
\end{equation}
In the presence of SU(2) symmetry,
the five spin-2 states generated by the above three basis vectors read
\begin{equation} 
\begin{aligned}
|2,\pm 2\rangle=|1,\pm 1\rangle|1,\pm 1\rangle,\\
|2,\pm 1\rangle=\frac{1}{\sqrt{2}}(|1,\pm 1\rangle|1,0\rangle+|1,0\rangle|1,\pm 1\rangle),\\
|2,0\rangle=\frac{1}{\sqrt{6}}(|1,1\rangle|1,-1\rangle+|1,-1\rangle|1,1\rangle+2|1,0\rangle|1,0\rangle).
\end{aligned}
\end{equation}
While for total spin angular momentum 1 and 0,
which require the formation of singlet,
the generated states are
\begin{equation} 
\begin{aligned}
|1,\pm 1\rangle=\frac{1}{\sqrt{2}}|1,\pm 1\rangle|1,0\rangle-|1,0\rangle|1,\pm 1\rangle,\\
|1,0\rangle=\frac{1}{\sqrt{2}}(|1, 1\rangle|1,-1\rangle-|1,-1\rangle|1, 1\rangle),\\
|0,0\rangle=\frac{1}{\sqrt{3}}(|1,1\rangle|1,-1\rangle+|1,-1\rangle|1,1\rangle-|1,0\rangle|1,0\rangle).
\end{aligned}
\end{equation}
Note that eigenstates $|S,m_{y}\rangle$ with same $m_{y}$ and different $S$ (eigenvalue) are mutually orthogonal,
i.e.,
$\langle S,m_{y}|S',m_{y}\rangle\sim \delta_{S,S'}$.

When the pairing of excited states in Cooper channel (triplet state insteads of singlet) with total spin 2 emerge
in the bulk,
they give rise to the nonlocal effect and may destroy the two-fold degeneracy in entanglement spectrum.
We write these five linear independent but not mutually orthogonal degeneracy states (with SU(2) in this subspace)
as 
\begin{equation} 
\begin{aligned}
|S,j,m_{y}\rangle=|2,j,m_{y}\rangle=(-1)^{j}(S^{-})^{m}(S^{+})^{2}|0\rangle, (m=0,1,2,3,4),
\end{aligned}
\end{equation}
where $|0\rangle$ denotes the ground state,
$j$ denotes the position of the operator in the chain
which results in the spin-2 degenerate multiplet (SU(2)),
and we here use the notation of following spin-1 operators ($+,-,z$ only denote the operator indices but no the actual spin directions)
\begin{equation} 
\begin{aligned}
S^{+}=\begin{pmatrix}
0 & 1 & 0\\
0 & 0 & 1\\
0 & 0 & 0
\end{pmatrix},\\
S^{-}=\begin{pmatrix}
0 & 0& 0\\
1& 0 & 0\\
0 &1 & 0
\end{pmatrix},\\
S^{z}=\begin{pmatrix}
1 & 0 & 0\\
0 & 0 & 0\\
0 & 0 & -1
\end{pmatrix}.
\end{aligned}
\end{equation}
The term $(S^{-})^{m}(S^{+})^{2}$ within the five degenerate excited states reads
\begin{equation} 
\begin{aligned}
&m=0:
(S^{+})^{2}=\begin{pmatrix}
0 & 0 & 1\\
0 & 0 & 0\\
0 & 0 & 0
\end{pmatrix},\\
&m=1:
[S^{-},(S^{+})^{2}]=[S^{-},S^{+}]S^{+}+S^{+}[S^{-},S^{+}]=-S^{z}S^{+}-S^{+}S^{z}=-\{S^{z},S^{+}\}\\
&=
\begin{pmatrix}
0 & -1& 0\\
0& 0 & 1\\
0 &0 & 0
\end{pmatrix},\\
&m=2:
[S^{-},-\{S^{z},S^{+}\}]
=\begin{pmatrix}
1 & 0 & 0\\
0 & -2 & 0\\
0 & 0 & 1
\end{pmatrix},\\
&m=3:
\begin{pmatrix}
0 & 0 & 0\\
1 & 0 & 0\\
0 & -1 &0
\end{pmatrix},\\
&m=4:
\begin{pmatrix}
0 & 0 & 0\\
0 & 0& 0\\
1& 0 & 0
\end{pmatrix}.
\end{aligned}
\end{equation}

\end{large}
\renewcommand\refname{References}

\clearpage

\clearpage


\begin{thebibliography}{99}
\bibitem{Bi Z}Bi Z, Zhang R, You Y Z, et al. Bilayer graphene as a platform for bosonic symmetry-protected topological states[J]. Physical review letters, 2017, 118(12): 126801.
\bibitem{Jian C M}Jian C M, Xu C. Generic “unnecessary” quantum critical points with minimal degrees of freedom[J]. Physical Review B, 2020, 101(3): 035118.
\bibitem{Vadimov V}Vadimov V, Hyart T, Lado J L, et al. Many-body Majorana-like zero modes without gauge symmetry breaking[J]. Physical Review Research, 2021, 3(2): 023002.
\bibitem{Fendley P}Fendley P. Parafermionic edge zero modes in Zn-invariant spin chains[J]. Journal of Statistical Mechanics: Theory and Experiment, 2012, 2012(11): P11020.
\bibitem{Moudgalya S}Moudgalya S, O Brien E, Bernevig B A, et al. Large classes of quantum scarred Hamiltonians from matrix product states[J]. Physical Review B, 2020, 102(8): 085120.
\bibitem{Fidkowski L}Fidkowski L, Kitaev A. Topological phases of fermions in one dimension[J]. Physical review b, 2011, 83(7): 075103.
\bibitem{Huse D A}Huse D A, Nandkishore R, Oganesyan V, et al. Localization-protected quantum order[J]. Physical Review B, 2013, 88(1): 014206.

\bibitem{Moudgalya S2}Moudgalya S, Rachel S, Bernevig B A, et al. Exact excited states of nonintegrable models[J]. Physical Review B, 2018, 98(23): 235155.

\bibitem{Shiraishi N}Shiraishi N, Mori T. Systematic construction of counterexamples to the eigenstate thermalization hypothesis[J]. Physical review letters, 2017, 119(3): 030601.
\bibitem{Hermele M}Hermele M, Gurarie V, Rey A M. Mott insulators of ultracold fermionic alkaline earth atoms: Underconstrained magnetism and chiral spin liquid[J]. Physical Review Letters, 2009, 103(13): 135301.
\bibitem{Chandran A}Chandran A, Khemani V, Laumann C R, et al. Many-body localization and symmetry-protected topological order[J]. Physical Review B, 2014, 89(14): 144201.

\bibitem{Moudgalya S3}Moudgalya S, Regnault N, Bernevig B A. Entanglement of exact excited states of Affleck-Kennedy-Lieb-Tasaki models: Exact results, many-body scars, and violation of the strong eigenstate thermalization hypothesis[J]. Physical Review B, 2018, 98(23): 235156.
\bibitem{Schecter M}Schecter M, Iadecola T, Sarma S D. Configuration-controlled many-body localization and the mobility emulsion[J]. Physical Review B, 2018, 98(17): 174201.

\bibitem{Garratt S J}Garratt S J, Chalker J T. Local pairing of Feynman histories in many-body Floquet models[J]. Physical Review X, 2021, 11(2): 021051.
\bibitem{Garratt S J2}Garratt S J, Chalker J T. Many-body delocalization as symmetry breaking[J]. Physical Review Letters, 2021, 127(2): 026802.
\bibitem{Brouwer P W}Brouwer P W, Beenakker C W J. Diagrammatic method of integration over the unitary group, with applications to quantum transport in mesoscopic systems[J]. Journal of Mathematical Physics, 1996, 37(10): 4904-4934.

\bibitem{O Brien E}O Brien E, Vernier E, Fendley P. “Not-A”, representation symmetry-protected topological, and Potts phases in an S 3-invariant chain[J]. Physical Review B, 2020, 101(23): 235108.

\bibitem{Vernier E}Vernier E, O Brien E, Fendley P. Onsager symmetries in-invariant clock models[J]. Journal of Statistical Mechanics: Theory and Experiment, 2019, 2019(4): 043107.
\bibitem{Harada K}Harada K, Kawashima N, Troyer M. Néel and Spin-Peierls Ground States of Two-Dimensional S U (N) Quantum Antiferromagnets[J]. Physical review letters, 2003, 90(11): 117203.

\bibitem{Iyoda E}Iyoda E, Katsura H, Sagawa T. Effective dimension, level statistics, and integrability of Sachdev-Ye-Kitaev-like models[J]. Physical Review D, 2018, 98(8): 086020.
\bibitem{Orus R}Orus R. A practical introduction to tensor networks: Matrix product states and projected entangled pair states[J]. Annals of Physics, 2014, 349: 117-158.


\bibitem{Pehlivan Y}Pehlivan Y. Quantum Invariants of the Pairing Hamiltonian[J]. arXiv preprint arXiv:0806.1810, 2008.

\bibitem{Lian B}Lian B, Sondhi S L, Yang Z. The chiral SYK model[J]. Journal of High Energy Physics, 2019, 2019(9): 1-55.

\bibitem{Hu Y}Hu Y, Lian B. The Chiral Sachdev-Ye Model: Integrability and Chaos of Anyons in 1+ 1d[J]. arXiv preprint arXiv:2109.13263, 2021.
\bibitem{Nahum A}Nahum A, Ruhman J, Vijay S, et al. Quantum entanglement growth under random unitary dynamics[J]. Physical Review X, 2017, 7(3): 031016.
\bibitem{Konig E J}Konig E J, Ostrovsky P M, Protopopov I V, et al. Metal-insulator transition in two-dimensional random fermion systems of chiral symmetry classes[J]. Physical Review B, 2012, 85(19): 195130.
\bibitem{You Y Z}You Y Z, Ludwig A W W, Xu C. Sachdev-Ye-Kitaev model and thermalization on the boundary of many-body localized fermionic symmetry-protected topological states[J]. Physical Review B, 2017, 95(11): 115150.

\end{thebibliography}
\end{document}